\documentclass[fleqn,usenatbib]{mnras}
\usepackage{times,txfonts}
\usepackage[T1]{fontenc}
\usepackage{ae,aecompl}
\usepackage{graphicx} 
\usepackage{pdflscape}

\def\fig{Figure}
\def\figs{Figures.}
\def\Fig{Figure}
\def\Figs{Figures}
\def\sect{Section}

\def\Sect{Section}

\def\tab{Table}
\def\tabs{Tables}

\def\Tabs{Tables}
\def\eqn{equation}

\newcommand{\object}[1]{#1}

\date{Accepted XXX. Received YYY; in original form ZZZ}

\pubyear{2017}

\title[IPS with the MWA I]{Interplanetary Scintillation with the Murchison Widefield Array I: A sub-arcsecond Survey over 900 square degrees at 79 and 158 MHz}
\author[J. S. Morgan et al.]{J .S. Morgan,$^{1}$\thanks{\href{mailto:john.morgan@icrar.org}{john.morgan@icrar.org}} J-P. Macquart,$^{1,2}$ R. Ekers,$^{3}$ R. Chhetri,$^{1,2}$\newauthor M. Tokumaru,$^{4}$ P. K. Manoharan,$^{5}$ S. Tremblay,$^{1,2}$ M. M. Bisi,$^{6}$ and B. V. Jackson$^{7}$ \\
$^{1}$International Center for Radio Astronomy Research, Curtin University, GPO Box U1987, Perth, WA 6845, Australia\\
$^{2}$ARC Centre of Excellence for All-Sky Astrophysics (CAASTRO)\\
$^{3}$CSIRO Astronomy and Space Science (CASS), P.O. Box 76, Epping, NSW 1710, Australia\\
$^{4}$Institute for Space-Earth Environmental Research, Nagoya University, Nagoya 464-8601, Japan\\
$^{5}$Radio Astronomy Centre, National Centre for Radio Astrophysics, Tata Institute of Fundamental Research, Ooty 643001, India\\
$^{6}$RAL Space, Science \& Technology Facilities Council, Rutherford Appleton Laboratory, Harwell Campus, Oxfordshire, OX11 0QX, UK\\
$^{7}$Center for Astrophysics and Space Sciences, University of California at San Diego, USA
}

\begin{document}
\label{firstpage}
\pagerange{\pageref{firstpage}--\pageref{lastpage}}
\maketitle
\begin{abstract}
We present the first dedicated observations of Interplanetary Scintillation (IPS) with the Murchison Widefield Array (MWA).
We have developed a synthesis imaging technique, tailored to the properties of modern ``large-N'' low-frequency radio telescopes.
This allows us to image the variability on IPS timescales across 900 square degrees simultaneously.
We show that for our observations, a sampling rate of just 2\,Hz is sufficient to resolve the IPS signature of most sources.
We develop tests to ensure that IPS variability is separated from ionospheric or instrumental variability.
We validate our results by comparison with existing catalogues of IPS sources, and near-contemporaneous observations by other IPS facilities.
Using just five minutes of data, we produce catalogues at both 79\,MHz and 158\,MHz, each containing over 350 scintillating sources.
At the field centre we detect approximately one scintillating source per square degree, with a minimum scintillating flux density at 158\,MHz of 110\,mJy, corresponding to a compact flux density of approximately 400\,mJy.
Each of these sources is a known radio source, however only a minority were previously known to contain sub-arcsecond components.
We discuss our findings and the prospects they hold for future astrophysical and heliospheric studies.
\end{abstract}
\section{Introduction}
Interplanetary Scintillation (IPS) is the phenomenon of brightness fluctuations in radio sources on $\sim$1\,s timescales due to scintillation induced by the turbulent solar wind plasma.
It was discovered by \citet{Clarke:phdthesis} who identified that only compact ($<$2\arcsec) sources showed these rapid scintillations.
Given that her observations also implied turbulent scales of 1\,km or more, this meant that the scattering screen must be much more distant than the ionosphere (the solar corona was noted as a plausible candidate).

\citet{1964Natur.203.1214H} realized the enormous astrophysical utility of a technique that can be used to identify radio sources that have sub-arcsecond components,
at a time before radio interferometers had sufficient resolution \citep{1962MNRAS.124..477A,2015IAUS..313..183H}.
This motivated the construction of the Cambridge IPS array, consisting of a large phased array with beams along the meridian, allowing the sky to be surveyed once a day at high time resolution. 
Most famously this instrument was used by Jocelyn Bell to discover pulsars \citep{1968Natur.217..709H}.
However it continued to be used for IPS for many decades, with almost 2000 IPS sources catalogued \citep{1987MNRAS.229..589P}.

Soon after the discovery of IPS, radio interferometry reached subarcsecond resolution \citep{1965Natur.205..375A}, and by the end of the 1960s VLBI observations had achieved milliarcsecond resolution \citep{2004evn..conf..237A}.
These advances were followed later by radio interferometers which had full imaging capabilities at GHz frequencies, where arcsecond resolution can be achieved with shorter baselines, and the effects of the ionosphere are reduced.
By the mid-1990s it became computationally feasible to perform all-sky surveys with these interferometers.
Most notably NVSS in the North \citep{1998AJ....115.1693C} and SUMSS in the South \citep{1999AJ....117.1578B}.

From the 1970s onward, heliospheric physics, rather than astrophysics, was the main driver for observations of IPS \citep[e.g.][]{1978SSRv...21..411C}.
Higher frequencies and multi-site observations allowed solar wind velocities and turbulence characteristics to be probed in regions where in-situ measurements had not yet been made: out of the ecliptic \citep{1967Natur.213..343D,1976JGR....81.4797C}, or close to the Sun \citep[e.g.][]{1971A&A....10..310E}.
Regular observations are currently conducted by a number of observatories, including the three-station ISEE array in Japan \citep{2011RaSc...46.0F02T}, Ooty in India \citep{1990MNRAS.244..691M}, and, more recently, MexART in Mexico \citep{2010SoPh..265..309M}.
These instruments (with the exception of Ooty), are dedicated to, IPS observations of $\sim$10--100 sources per day, which can then be used to reconstruct the 3-dimensional structure of the solar wind \citep{1998JGR...10312049J}\footnote{Daily maps of these reconstructions are produced, along with forecasts for the Earth's location. See \url{http://ips.ucsd.edu/}}

More recently, interest in low-frequency radio astronomy has been rekindled by Epoch of Reionisation studies.
This has motivated the construction of a new generation of radio telescopes including the Murchison Widefield Array \citep[MWA;][]{2013PASA...30....7T}, the LOw Frequency ARray \citep[LOFAR;][]{2013A&A...556A...2V}, and the Long Wavelength Array \citep[LWA;][]{2013ITAP...61.2540E}.
These instruments have been used for a range of science goals.
In particular, there has been a proliferation of all-sky surveys with these new instruments including 
the MWA GLEAM Survey \citep{2015PASA...32...25W,2017MNRAS.464.1146H},
and MSSS with LOFAR \citep{2015A&A...582A.123H},
as well as the alternative data release of the TGSS, conducted with the GMRT \citep{2017A&A...598A..78I},
and a redux \citep{2014MNRAS.440..327L} of VLSS \citep{2007AJ....134.1245C} conducted with the VLA.
In addition, IPS has already been detected and studied with both LOFAR \citep{2013SoPh..285..127F,2016ApJ...828L...7F} and the MWA \citep{2015ApJ...809L..12K}.

This latest generation of ``Large-N'' arrays \citep{2009IEEEP..97.1497L} have two properties which make them outstanding instruments for IPS studies.
The large number of array elements means that the instantaneous imaging fidelity is excellent.
Furthermore, since the individual elements are small, the field of view of the instrument is very large.

In this paper we introduce a novel method of making IPS measurements which makes use of these key properties of current and planned large-N arrays.
We show that it is possible to make IPS measurements simultaneously across the full field of view, allowing a census to be made of all compact emission across a wide area.
Moreover, we reintroduce IPS as a powerful astrophysical tool which provides a practical method for determining the arcsecond-scale structure of all sources across a large fraction of the sky, without the need for very long baselines and the calibration challenges they entail.

The article is laid out as follows: in \sect~\ref{sec:method} we review the methods by which a time series of flux density measurements can be obtained using an array.
We describe the power spectrum of an IPS signal and introduce the technique of variability imaging.
In \sect~\ref{sec:observations} we describe our observations and the data reduction process.
In \sect~\ref{sec:results} we describe the construction of a catalogue of scintillating sources from our observations, and in \sect~\ref{sec:discussion} we discuss the implications of these findings for future heliospheric and astrophysical research.
In the 2nd paper in this series \citep{2017Chhetri} IPS measurements with the MWA are synthesized with spectral information from GLEAM and other radio and gamma-ray surveys to understand the astrophysical nature of the low-frequency compact source population.
\section{Methods}
\label{sec:method}
\subsection{High time resolution flux density measurements}
A number of approaches can be taken to convert the voltages measured by each of the $N$ elements of an array and convert them into flux density measurements. 

\subsubsection{Beamforming}
One widely-used approach is to sum the voltage time series from each element with appropriate delays applied to each such that the time series are combined in-phase for a particular point on the sky.
Once combined the timeseries is then squared and integrated up to the desired time resolution.
The area on the sky to which the instrument is sensitive is now the synthesized beam of the entire array.
This approach is possible with the MWA using the voltage capture system \citep{2015PASA...32....5T}, and has been used in recent IPS studies using LOFAR.
The disadvantage of this approach is that a beam must be formed for each IPS source, and for the MWA there could be many hundreds of sources within the field of view.

\subsubsection{Correlation and Synthesis Imaging}
Radio interferometers such as the MWA and LOFAR are primarily designed and used as interferometers.
The voltage data from each pair of elements are correlated to form visibilities.
These are then gridded and inverted to form a 2D image of the sky, with the field of view limited only by the response of the individual elements.
In order for this technique to be applied to IPS observation, this must be repeated on a high enough time cadence to resolve IPS variability.
The result is an image ``cube'' with two celestial dimensions and a time dimension.

Such a cube could also be constructed by forming a very large number of beams, one per pixel of the image.
Indeed precisely that approach has been taken in order to facilitate high time resolution imaging of the Sun with LOFAR \citep{2014A&A...568A..67M}.
However, discrete Fourier transforms can be computed highly efficiently, which makes synthesis imaging vastly more computationally feasible for large fields.
This must be balanced against the necessity to form $N\times N$ pixels across the field of interest, only a tiny fraction of which will contain a source above the detection limit. 

Synthesis imaging has numerous advantages however.
Since the correlated power is calculated for each individual baseline, each can be weighted appropriately.
By including only cross-correlations between antennas (zero-weighting the autocorrelations), and down-weighting the shorter baselines, the signal-to-noise of compact sources can be boosted relative to extremely bright but extended sources such as the quiet Sun and diffuse Galactic emission.
This also reduces the impact of RFI, which is stronger in the autocorrelations and shorter baselines.
In contrast, weighting of data in coherent beamforming can only be done on an element-by-element basis and the power will be dominated by Galactic emission (at least at metre wavelengths).

Imaging the full field of view allows deconvolution to be used to improve image fidelity.
For IPS studies, this means mitigating the effect that a bright scintillator will have on neighbouring sources.

The image cube will contain time series not just for pixels containing scintillating sources, but for non-scintillating sources, as well as off-source pixels.
These can be used to give a highly detailed picture of the noise properties of the observation (see \sect~\ref{sec:var_imaging}).

Finally, if the \textit{a priori} location of a source is wrong (for example, because the apparent location of the source has been shifted by ionospheric refraction), then the source will still be captured, and there will be no ambiguity as to where the source centroid is, even if this changes during the observation.
IPS and ionospheric effects may be distinguished in the image plane (see \sect~\ref{sec:ionosphere_image})
Furthermore, the imaging technique is an effective way to discover new scintillating sources, and the imaging cubes might even be used to detect fast radio transients \citep[indeed the IPS reported by ][was discovered while searching for Fast Radio Bursts in MWA data]{2015ApJ...809L..12K}.

\subsubsection{Correlation and Non-imaging analysis of visibilities}
Finally, we note that the interferometry approach does not necessarily require images to be made.
For example, \citet{2008ISTSP...2..707M} have developed a system capable of monitoring the location and flux density of large numbers of sources in the visibility domain.
\citet{2017arXiv170704978J} have used this successfully for monitoring the locations of 1000 sources with an 8s time cadence, and similar techniques can be used for flux density measurements of known sources on IPS timescales \citep{2015ApJ...809L..12K}.
A hybrid approach is also possible, whereby known sources are characterised and subtracted from the visibilities, and the residuals are imaged.
\subsection{Time series analysis}
The fundamental observable in IPS is a timeseries of flux density measurements, preferably with sufficient time resolution to fully resolve IPS variability.
Convolved with any IPS variability are various other propagation and instrumental effects, which must be accounted for if the IPS signal is to be accurately recovered.
Fortunately these signals can easily be separated, since they manifest differently and on different timescales.

\subsubsection{Forming a power spectrum}
A number of methods exist for forming a power spectrum from a regularly spaced time series.
Many involve smoothing or averaging either before or after transforming to the spectral domain.
Welch's method \citep{welch:1967} is preferred in our case since no averaging is done, and we are interested in signals up to and including the Nyquist frequency (1\,Hz for our data).
In \figs~\ref{fig:ips_source} and \ref{fig:non_ips_source} we show power spectra generated with FFT sizes of 16 (to zoom in on the IPS band with minimum error bars) and 144 (to show greater resolution in the IPS band and show lower frequencies).
In both cases we use a Hann window function to reduce sidelobes in the power spectrum domain, and use overlapping windows to recover signal from samples at the edge of each window.
The error bars are slightly asymmetric since the power spectrum estimates of Gaussian white noise are $\chi^2$ distributed \citep{welch:1967} and show the 67\% confidence interval (i.e. approximately equivalent to 1-$\sigma$).
Note that each time sample is an integration over the full 0.5s (not an instantaneous measurement), so there is no danger of our power spectrum being corrupted by aliased power from above the Nyquist frequency.

\subsubsection{The IPS signal}
\label{sec:ips_signal}
\begin{figure*}
  \includegraphics[width=\textwidth]{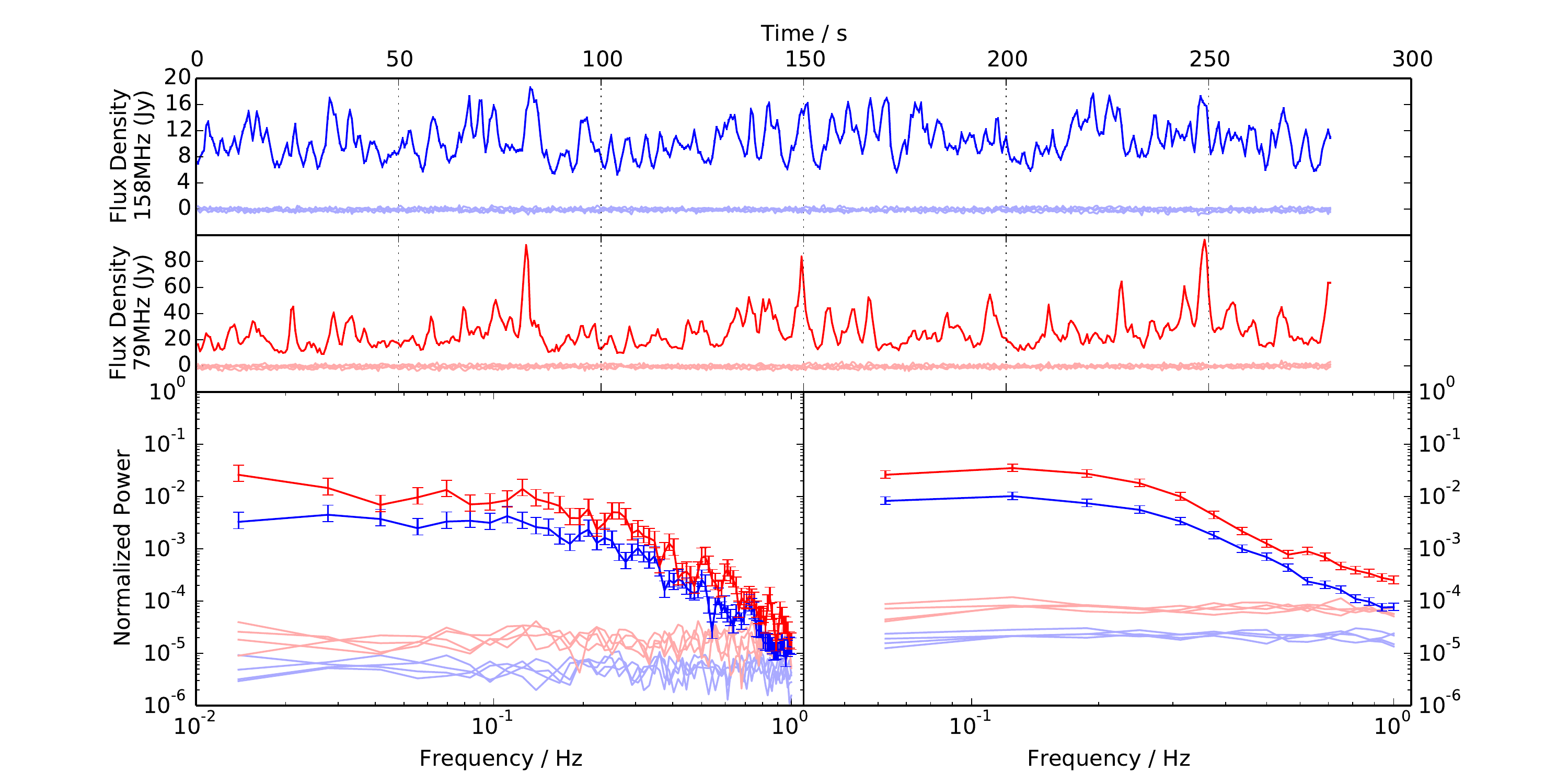}\\
  \caption{Time series and spectra for IPS source \object{3C2}. Red is used for low band, Blue for high band. Saturated tones are used for on-source measurements, pastel tones are used for off-source measurements from four nearby pixels. Top two panels are time series, bottom two panels show power spectra of the time series with different resolutions.}
  \label{fig:ips_source}
\end{figure*}
\begin{figure*}
  \includegraphics[width=\textwidth]{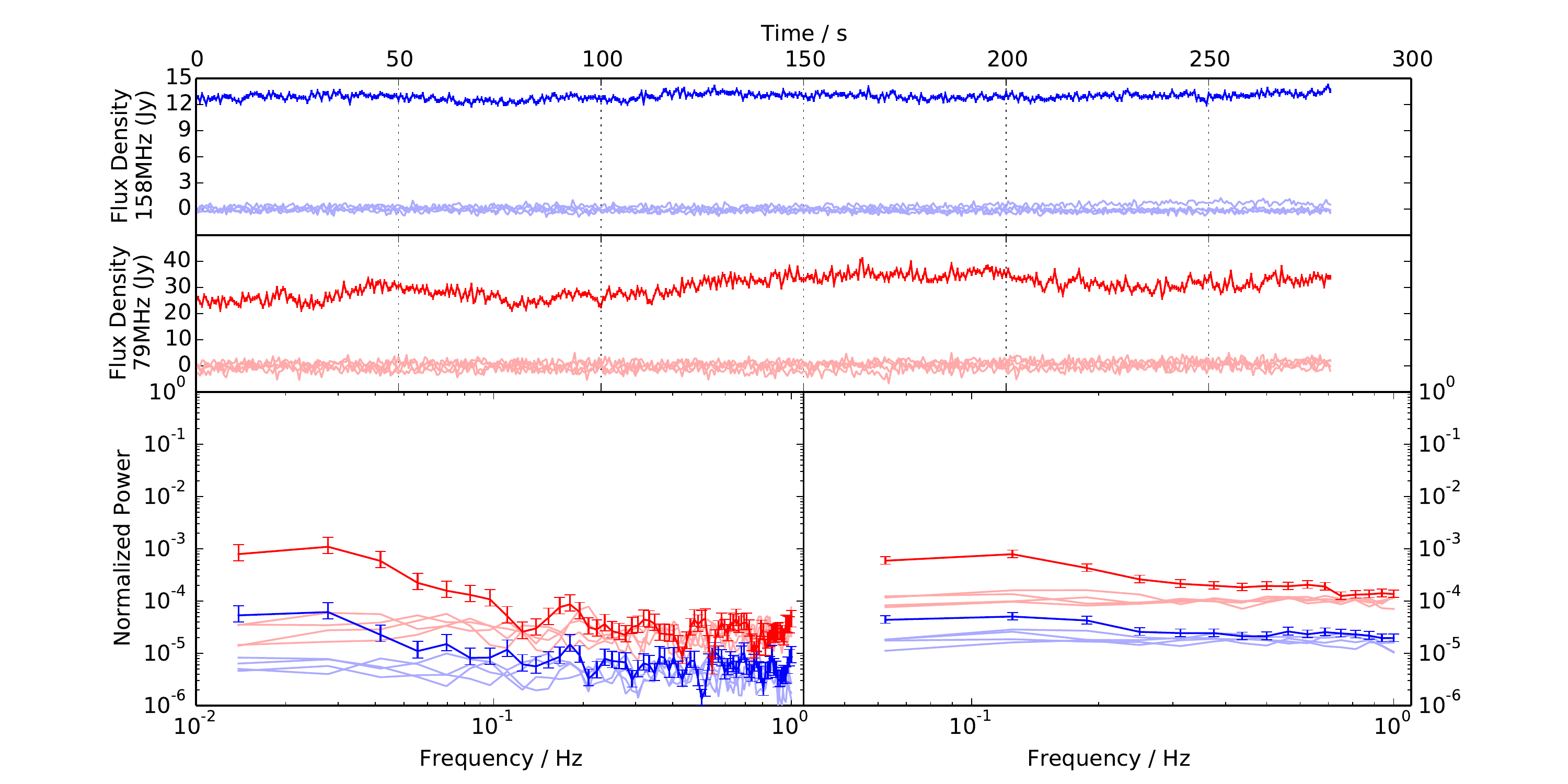}\\
  \caption{Time series and Power Spectra for source which shows little or no IPS (\object{3C18}). Red is used for low band, Blue for high band. Saturated tones are used for on-source measurements, pastel tones are used for off-source measurements from four nearby pixels. Top two panels are time series, bottom two panels show power spectra of the time series with different resolutions.}
  \label{fig:non_ips_source}
\end{figure*}
Time series and power spectra are shown in \fig~\ref{fig:ips_source} for a very bright source which shows clear IPS (3C2) and in \fig~\ref{fig:non_ips_source} for a very bright source which shows little or no IPS (3C18). 
Time series and corresponding power spectra for a number of nearby off-source pixels are also shown.
In the time domain these have a mean close to zero, and within the range 0.1--1Hz, their spectra are flat (i.e. white noise), and agree with each other within errors.

In contrast to the data used in many IPS studies, the mean flux density of the source is preserved as the mean of the on-source time series.
Thus, rather than arbitrarily scaling the power spectrum, we normalize it such that the integral of the power spectrum with respect to frequency gives the square of modulation index, $m$ (where $m$ is the ratio of the rms to the mean).
Note that the mean of the time series is only one of a number of ways to determine the mean flux density of the source, this is discussed further in \sect~\ref{sec:scint_index}.

The IPS signature in the range 0.1--1Hz is well fit by an empirical function of the form
\begin{equation}
	\frac{A}{1+\left(\frac{\nu}{\nu_0}\right)^\alpha}
	\label{eqn:butterworth}
\end{equation}
where the power asymptotically approaches $A$ towards low frequency, $\nu_0$ is the frequency of the half power point, and $\alpha$ is the index of the power law drop off at high frequency.
For the source shown in \fig~\ref{fig:ips_source}, $\nu_0$ is 0.3\,Hz (0.25\,Hz in the lower band), and $\alpha$ is approximately 4.5.
This is consistent with the findings of \citet{2013SoPh..285..127F} and \citep{2011RaSc...46.0F02T} at slightly higher frequencies, and \citet{2015ApJ...809L..12K} \citep[see also][]{2016ApJ...828L...7F} in their investigations of night-side IPS.

At the Nyquist frequency (1\,Hz), the on-source power is still slightly above the off source power. 
This means that the 0.5s time resolution is not quite sufficient to fully resolve the IPS variability, however by 1\,Hz the scintillating power has dropped more than 20\,dB below its peak value.
The vast majority of sources do not have this signal-to-noise and so the IPS is effectively resolved.
It is worth noting that if we are generating our timeseries via synthesis imaging, keeping the sampling time as long as possible will reduce number of images that have to be made, in turn reducing the computational cost.
Since the noise can be fully characterized using the off-source pixels, it is not necessary to preserve the high-frequency tail of the power spectrum in order to determine the noise level.
Therefore the sampling rate of 2\,Hz is close to optimum for this dataset.

\subsubsection{Other sources of variability}
Ionospheric scintillation differs from IPS in that it has a longer timescale (10\,s vs 1\,s), and affects sources up to a much larger size (10\arcmin\ vs $<$1\arcsec, see \citealp{Thompson:2001}).
This is considerably larger than all of the sources in our field, so we would expect all to be affected by ionospheric scintillation.

Variability on timescales $>$10\,s is clearly visible in \fig~\ref{fig:non_ips_source}, both in the time series and the power spectrum, particularly in the low band.
However, almost all of this variability is below 0.1\,Hz, meaning that ionospheric scintillation can be filtered out almost entirely by discarding just the lowest 10\% of the power spectrum.
Some variability does remain above 0.1\,Hz, however whether this is IPS, ionospheric scintillation, or some other effect is unclear.
In any case it represents a variability of $<1$\%, which would not be detectable for any but the very brightest sources.
In \sect~\ref{sec:ionosphere_image} we present further evidence in favour of this being low-level of IPS.

Spurious variability may also be caused due to the instrumental response to bright sources across the field of view.
The very large number of elements in the MWA and their pseudorandom arrangement mean that its Point Spread Function (PSF) is exceptionally good.
However, on the very longest baselines, the synthesized aperture is far from being filled, causing sidelobes around each source.
As the Earth rotates, these sidelobes rotate, sweeping over the rest of the image, causing spurious variability.
Even assuming sidelobes have the size of the maximum spatial resolution of the MWA ($\sim$ 4\arcmin) and that the source is in a distant \textit{tile} sidelobe ($\sim$100\degr\ away), the characteristic frequency of this variability would still be 0.05Hz.
However there is no evidence of excess power at 0.05Hz and below in the off-source power spectra in \figs~\ref{fig:ips_source} \& \ref{fig:non_ips_source}, and even if there were, the low frequency of the variability and the fact that it affects both on- and off-source parts of the image would make it easily separable from IPS.
This effect may complicate ionospheric scintillation studies with the MWA, though it should be quantifiable using off-source variability.

\subsection{Variability Imaging}
\label{sec:var_imaging}
In the previous section we set out how we can efficiently produce a timeseries for every point within the field of view of the MWA.
We have also characterized IPS in the time domain, and shown how it can be separated from other sources of variability.
We now wish to construct a summary statistic of a time series, which will allow us to collapse the time dimension of our image cube, and produce an image which shows the scintillating flux density for each pixel of our image, while filtering out any non-IPS sources of variability.

\subsubsection{Time domain filtering}
In order to remove non-IPS signals, it is necessary to remove variability on timescales of $>$10\,s.
In addition, since the majority of IPS variability is concentrated in the lower half of our power spectrum, we can improve our signal to noise, by down-weighting variability on timescales $<1$\,s.

Conveniently, the Low-pass Butterworth filter has precisely the same functional form as our empirical characterization of IPS (see \eqn~\ref{eqn:butterworth}).
Thus by passing our timeseries through a Butterworth bandpass filter, which attenuates variability below 0.1Hz, and above 0.5Hz, we produce a timeseries which contains only the desired IPS signal, along with some (white, Gaussian) system noise.

\subsubsection{The variability image}
Recall that the output of our synthesis imaging process is an image ``cube'' with two spatial dimensions and a time dimension.
We can generate an image of the variability on IPS timescales by taking the standard deviation of the filtered time series.
Since the filtered time series has zero mean (the filtering has subtracted the DC component), the standard deviation is equivalent to the root-mean-square (RMS).
We term this image the ``variability image'' to avoid confusion with images of the \emph{spatial} RMS of a synthesis image.

We now determine the properties of this image.
Consider the signal-absent case (no scintillating source present) for an image generated from time series of $N$ points.
The filtered time series values will be drawn from a Gaussian distribution of zero mean and standard deviation equal to the system noise of a single time integration, at that location on the sky.
Thus for sufficiently large $N$, the pixel values $P_{x,y}$ in the variability image will be drawn from a Gaussian distribution of mean $\mu$ (where $\mu$ is equal to the system noise) and standard deviation $\sigma$ where
\begin{equation}
	\sigma = \frac{\mu}{\sqrt{N}}.	
	\label{eqn:sigma}
\end{equation}
Any scintillating sources in the image will give rise to a pixel value in excess of that due to the system noise.
We can then identify all scintillating sources by identifing those pixels in the variability image where
\begin{equation}
	P_{x,y}-\mu > 5\sigma.
	\label{eqn:detection_threshold}
\end{equation}
Both $\mu$ and $\sigma$ can be estimated for any location in the image from the distribution of the pixel values in the area of interest.
It is better to estimate $\sigma$ directly from the distribution of pixel values, rather than inferring it from \eqn~\ref{eqn:sigma} since the former will take account of any other sources of noise.

These Gaussian statistics are very convenient, since standard source-finding tools, such as Aegean \citep{2012MNRAS.422.1812H}, which uses a scheme very similar to the one outlined, can be used to detect and characterize the sources in the variability image.
These tools will also fit an elliptical Gaussian to the pixel values, eliminating any error due to pixellation, and giving a non-quantised estimate of the location and value of the peak $P$ in variability.

\subsubsection{Deriving the scintillating flux density}
\label{sec:scint_flux}
$P$ is a useful statistic for source finding, however it is \emph{not} the same as the scintillating flux density, $\Delta S_{scint}$, which, if we assume all excess variance is due to IPS, is given by
\begin{equation}
	\Delta S_{scint} = \sqrt{P^2 -\mu^2} .
	\label{eqn:delta_s_s}
\end{equation}
In the high S/N limit($P\gg\mu$), $\Delta S_{scint}\approx P-\mu$, and the uncertainty on $\Delta S_{scint}$ due to the system noise is simply $\sigma$.
However this will not be the case for the weakest sources if $N$ is reasonably large.
Setting a detection limit in the variability image of $P-\mu=5\sigma$, we find that when $P-\mu \ll \mu$, the weakest detectable scintillating flux density is
\begin{equation}
	\Delta S_{5\sigma} \approx \mu \sqrt{\frac{10}{\sqrt{N}}} .
	\label{eqn:delta_s_min}
\end{equation}
In other words the detection limit only decreases as the \emph{fourth} root of the observing time.

We take the error on the scintillating flux density due to system noise to be 
\begin{equation}
	\sigma_{sys}\left(\Delta S\right) = \sqrt{\left(P+\sigma\right)^2 - \mu^2}-\Delta S_{scint},
	\label{eqn:delta_s_sys}
\end{equation}
neglecting the slight asymmetry in the positive and negative errors.
We must then account for the error on the measurement of the scintillating flux density due the finite number of samples over which we measure it. 
The number of independent time samples depends on the timescale of the IPS signal in samples, $n_{IPS}$.
We add this in quadrature to $\sigma_{sys}\left(\Delta S\right)$ so that the error on $\Delta S_{scint}$ is

\begin{equation}
	\sigma\left(\Delta S\right)  = \sqrt{\left(\sigma_{sys}\left(\Delta S\right)\right)^2 + \frac{\Delta S_{scint}^2}{N/n_{IPS}}}.
	\label{eqn:err_delta_s}
\end{equation}
We take $n_{IPS}$ to be 3 (1.5\,s) in the high band and 4 (2\,s) in the low band (see \sect~\ref{sec:ips_signal}).
\subsubsection{Higher moments}
Other summary statistics may be useful in determining the presence and nature of IPS.
As shown in \fig~\ref{fig:ips_source}, the power spectrum of the IPS is almost identical for the low and high bands, however inspection of the time series reveals differences.
The fluctuations in the high band are Gaussian in distribution and easily temporally resolved.
The Gaussian distribution of flux densities is to be expected in the weak scintillation regime.
The fluctuations in the low band are more typical of strong scintillation, being exponential in distribution, with sharp peaks (i.e. not temporally resolved).
Higher order moments (e.g. skew, kurtosis), should be a good discriminator of these characteristics.
However any smoothing of the time series (e.g. by a low-pass filter) will regress the timeseries towards a Gaussian distribution, and blunt these statistics. 

Therefore the summary statistics should be calculated from the time series as follows:-
\begin{enumerate}
	\item Calculate mean;
	\item Apply high-pass filter;
	\item Calculate higher-order moments (skew, kurtosis);
	\item Apply low-pass filter;
	\item Calculate RMS (for variability image).
\end{enumerate}

\section{Observations}
\label{sec:observations}
\begin{table}
  \footnotesize
  \centering
  \caption{\label{tab:observations}Observation, correlation and imaging parameters.}
  \begin{tabular}{lcc}
    \hline
                          & Low band                                   & High Band                                \\
    \hline                                                                                                                           
    Central Frequency     & 78.7\,MHz                                  & 158\,MHz                                 \\
    Bandwidth             & 12.8\,MHz                                  & 15.36\,MHz                               \\
    Image Centre (J2000)  & 00:14:43 +11:44:18                         & 00:14:43 +11:44:18                       \\
    Primary Beam Centre   & 23:48:00 -00:30:00                         & 00:04:00 +08:15:00                       \\
    Half-power beam       & 1150 square degrees                        & 450 square degrees                       \\
    Image dimensions      & 2048$\times$2048                           & 2400$\times$2400                         \\
    Pixel size            & 2\arcmin$\times$2\arcmin                   & 1\arcmin$\times$1\arcmin                 \\
    PSF (at image centre) & 6.2\arcmin$\times$5.0\arcmin, PA 27\degr   & 3.6\arcmin$\times$2.7\arcmin, PA 31\degr \\
    \hline
  \end{tabular}
\end{table}
In order to allow multi-frequency IPS analysis it was decided to split the MWA's 30.72\,MHz of instantaneous bandwidth into two bands, spaced approximately a factor of two apart in frequency.
The observing parameters for both bands are summarized in \tab~\ref{tab:observations}.
Within the MWA observing band there is 15MHz of bandwidth directly below the FM broadcast band centred on 79\,MHz\footnote{While RFI levels at the MWA are substantially lower than less remote sites \citep{2015PASA...32....8O}, interference from FM stations is not unknown and the daytime RFI environment is less well characterized than the nighttime environment}.
An observing frequency of 158MHz, double the lower-band frequency, is close to the centre of the MWA's usable band and gives a good trade off between sensitivity, angular resolution and field of view.

To facilitate near simultaneous observations with the IPS array run by ISEE \citep{2011RaSc...46.0F02T}, a field was chosen centred on a relatively northern declination, with an Hour Angle of approximately 20\degr\ to match the ISEE array meridian.

The presence of the Sun in the field makes it extremely difficult to see any other radio source.
It is therefore necessary to ensure that the Sun is very highly attenuated by the primary beam.
This is not achievable for both bands at once, since the location of the nulls is frequency dependent.
However the quiet Sun has a spectral index of $+2.0$ at metric wavelengths, whereas astrophysical sources have a spectral index $\sim -0.7$, so the brightness of the Sun relative to the astrophysical sources we are observing is almost an order of magnitude greater in the higher band.
We therefore aimed to place the Sun in the null in the high band.

Two hours of observations were taken on 2015-04-29, with the array pointed statically at an azimuth of 31\degr\ and elevation 46\degr.
This corresponds to a declination of +12\degr.
Observations ran from 00:37UTC to 02:26UTC during which time the Sun moved from 32\degr\ to 6\degr\ from the field centre.
Test imaging of the data revealed that at the very start of the observation the Sun was almost perfectly nulled in the high band, and sufficiently attenuated in the low band that the apparent surface brightness of the Sun was comparable with the brightest astrophysical sources in the field (suggesting that the Sun is attenuated to just a few percent of its true flux density).
This first dataset, consisting of 560 timesteps (just under 5 minutes) was chosen for further analysis.

\subsection{Correlation and Pre-processing}
\label{sec:preprocessing}
These observations were correlated online using the standard MWA correlator \citep{2015PASA...32....6O} with the maximum time resolution available (0.5s time resolution, 40kHz spectral resolution).
Cotter \citep{2015PASA...32....8O} was used to flag the data for RFI, and apply instrumental and geometric delays to produce a set of visibilities with the phase centre set to the pointing centre of the array.
Standard MWA calibration techniques \citep{2014PASA...31...45H} were used to derive a calibration solution from an observation of 3C444 taken just before the sunrise following the IPS observations.
The calibration solutions for the lower band were unsatisfactory, so a solution was derived from an observation from the end of the night preceding the IPS observations.
Unfortunately, due to the observation having a slightly different observing frequency, 2.56\,MHz at the bottom of the low band had to be discarded.

\subsection{Standard Imaging}
\label{sec:standard_image}
These calibration solutions were then applied to the IPS observations, and WSClean \citep{2014MNRAS.444..606O} was used to make a synthesis image for each band, using the full five minutes of observing time, covering all of the sky where our sensitivity was at least 30\% that at the beam centre.
The two orthogonal linear polarizations were imaged separately (though WSClean's joint deconvolution scheme was used), and then combined using the primary beam model to weight each polarization appropriately.
This resulted in fairly typical MWA snapshot images. The Aegean source finding tool \citep{2012MNRAS.422.1812H} was used to locate $\sim$1000 sources above 5-$\sigma$ for each frequency.

\subsubsection{Correcting for refractive ionospheric effects}
\begin{figure*}
  \includegraphics[width=0.48\textwidth]{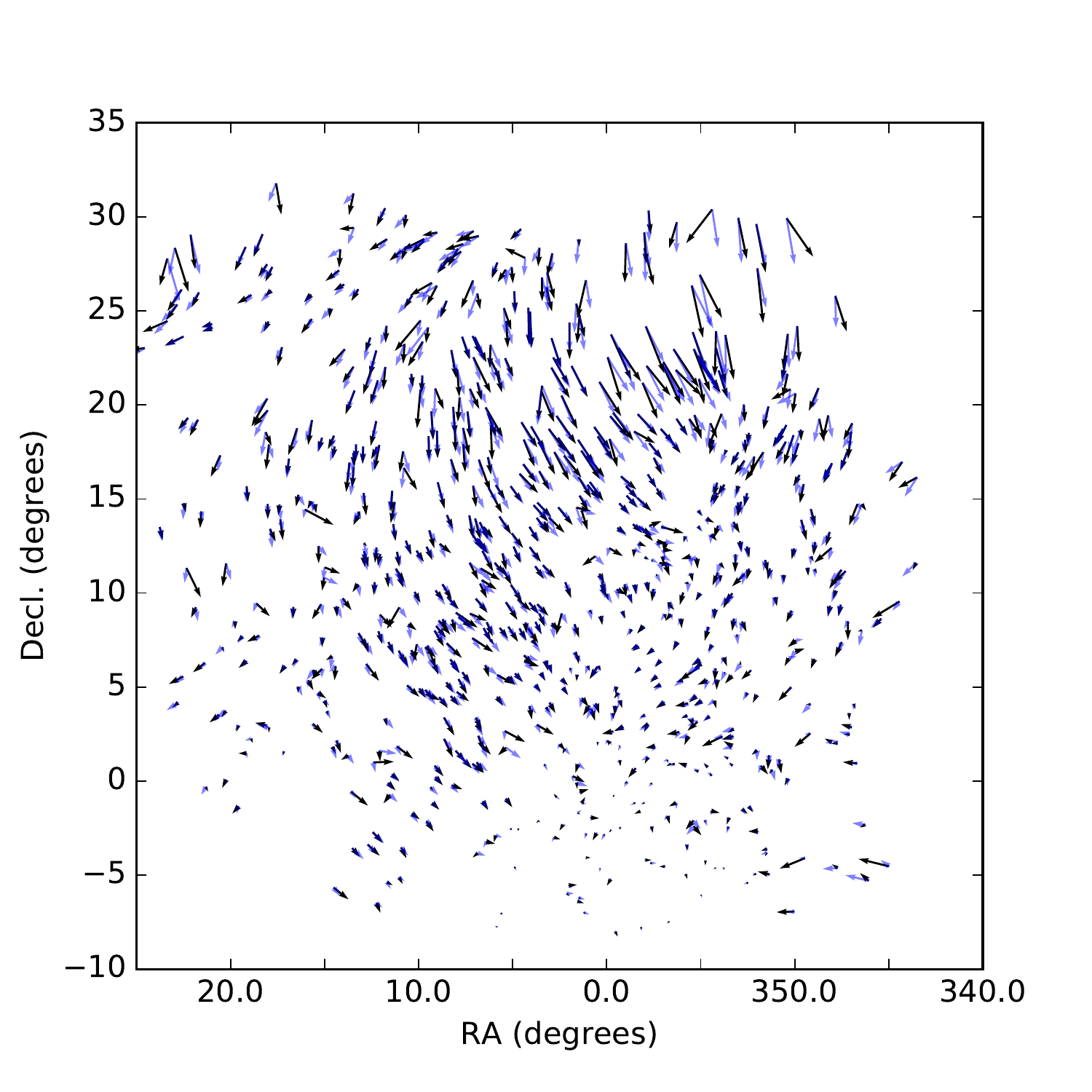}\,
  \includegraphics[width=0.48\textwidth]{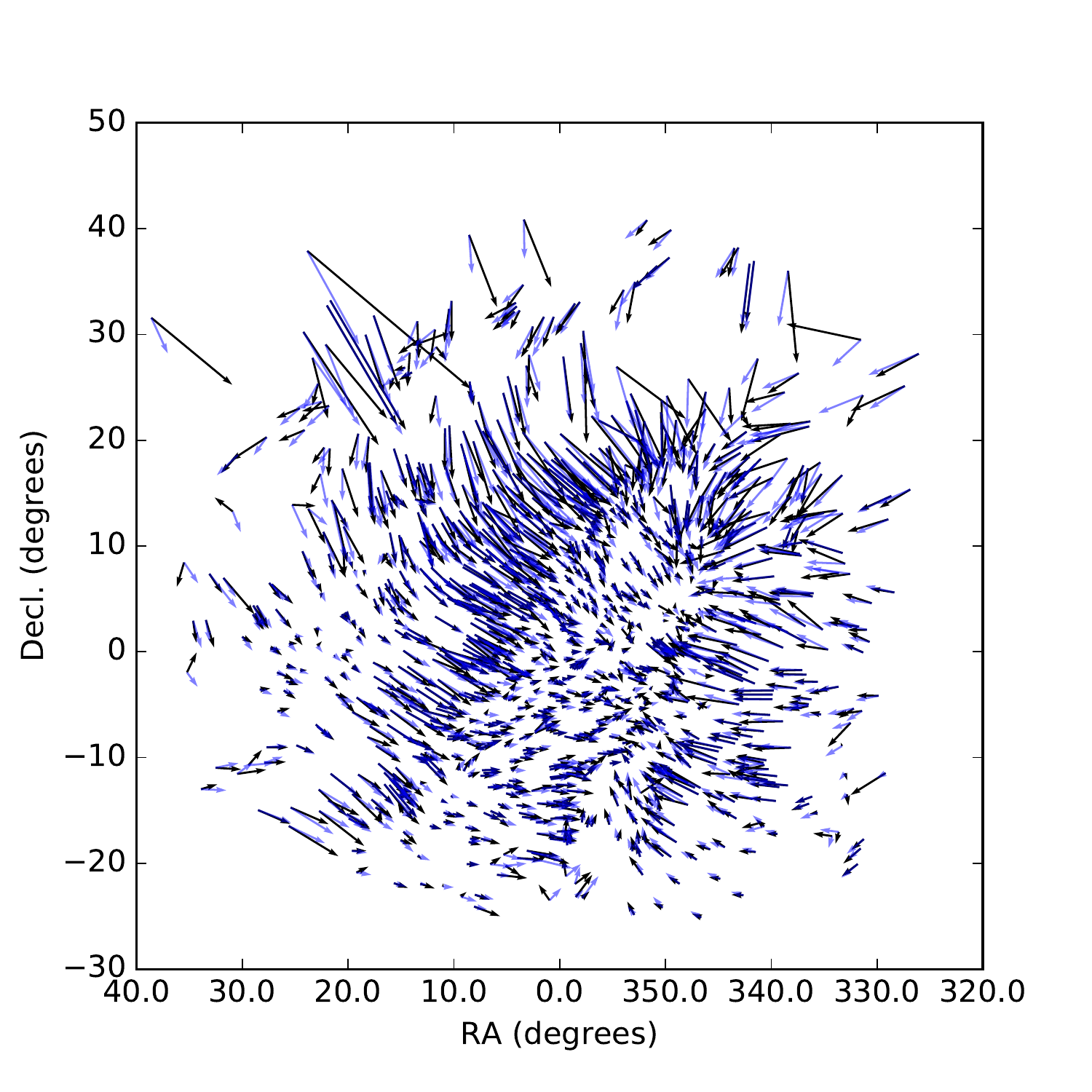}
  \caption{Plot of offsets of source apparent positions from their catalogued positions. High band is on the left, Low band is on the right. For lengths of arrows, $1\degr=1\arcmin$. Black arrows: Offset between apparent source position in our standard image and catalogued position; Blue arrows: position predicted from RBF using all sources except the source in question.}
  \label{fig:ionosphere_map}
\end{figure*}

\begin{figure}
    \includegraphics[width=0.23\textwidth]{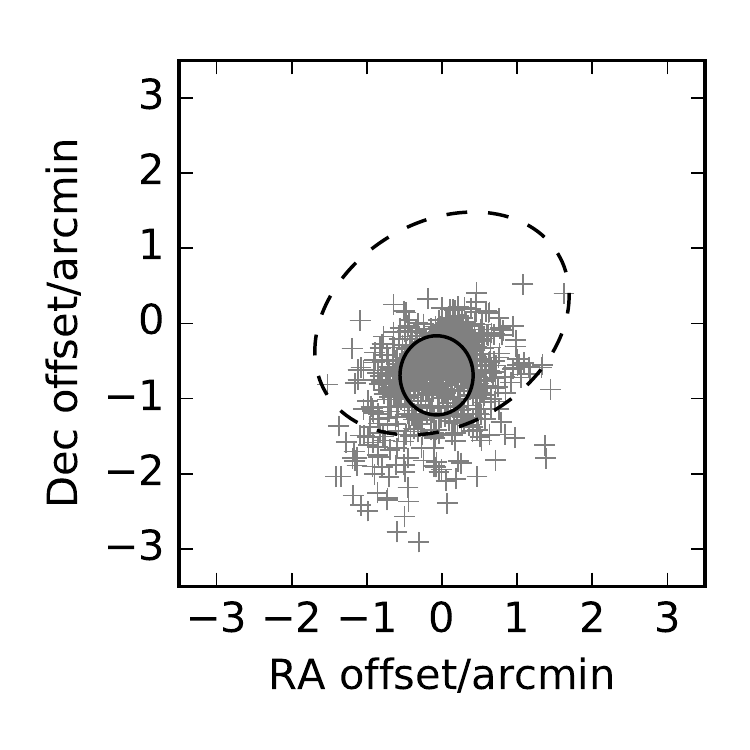}\,
    \includegraphics[width=0.23\textwidth]{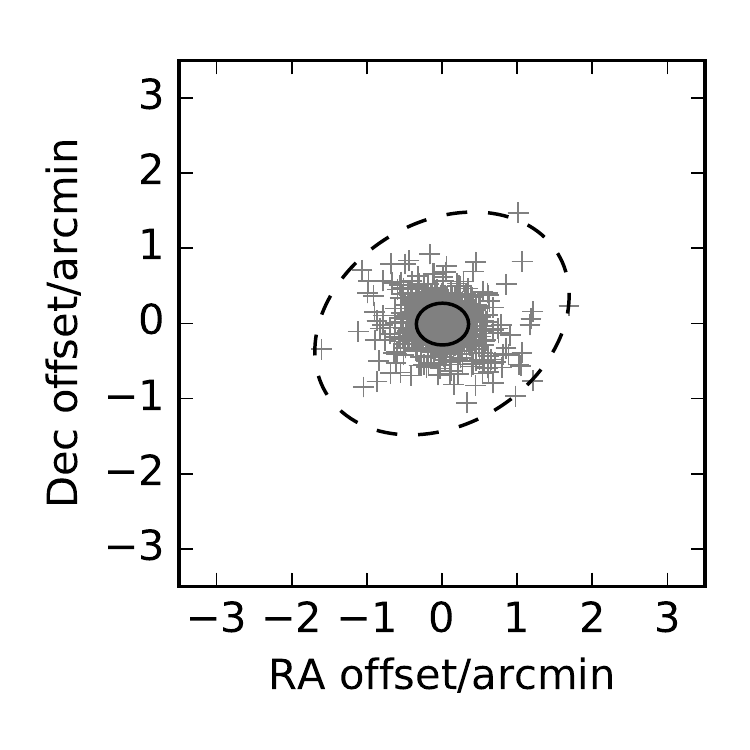}\\
    \includegraphics[width=0.23\textwidth]{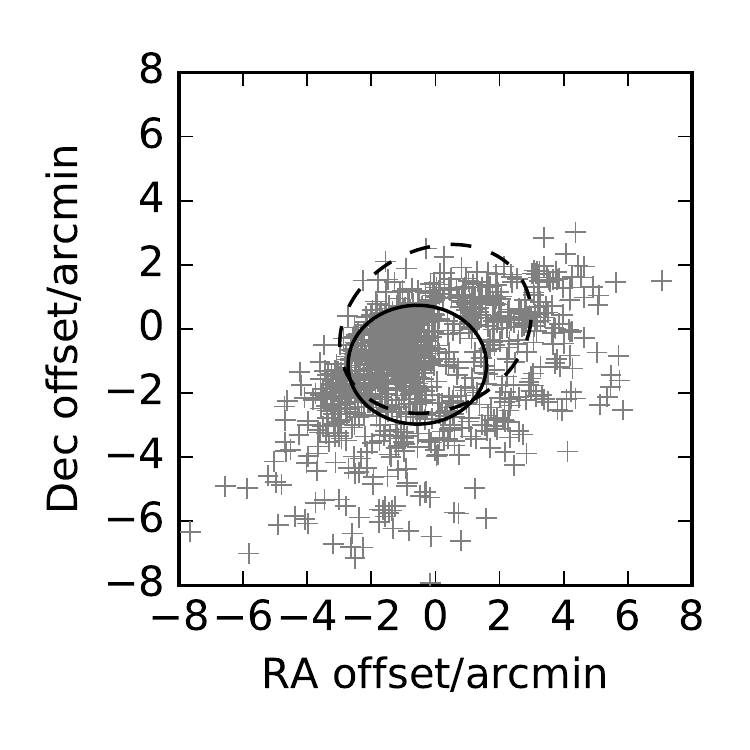}\,
    \includegraphics[width=0.23\textwidth]{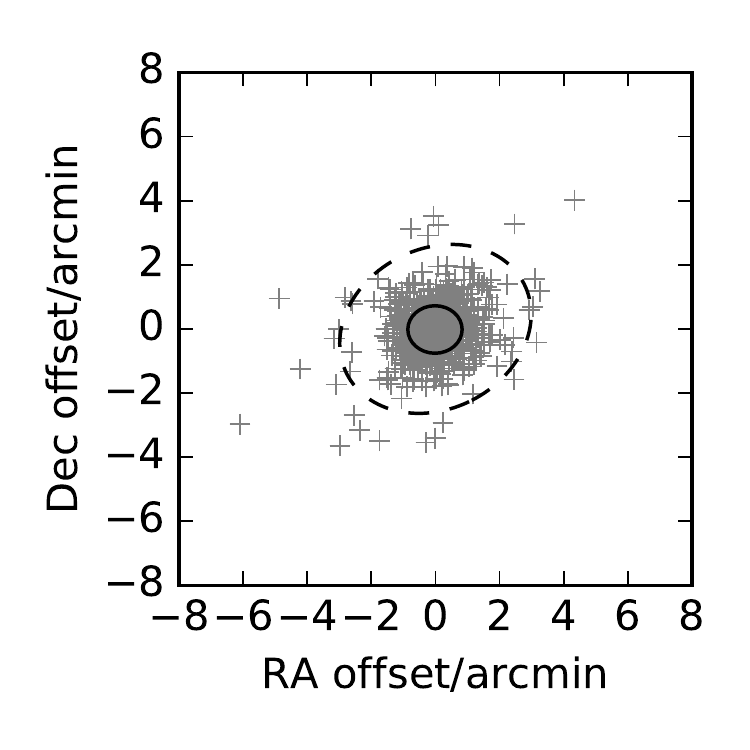}
    \caption{\label{fig:ionosphere_offsets}Left panels: offsets between source positions in our image vs VLSSr (i.e. Black arrows shown in \fig~\ref{fig:ionosphere_map}). Right panels: plot of vector difference of black arrows and blue arrows shown in \fig~\ref{fig:ionosphere_map}. Top: high band. Bottom: low band. mean and standard deviation are shown by location and size of solid ellipse. PSF is shown by dashed ellipse.}
\end{figure}

Next, the sources detected in the standard image were matched against the VLSSr catalogue \citep{2014MNRAS.440..327L}.
\Fig~\ref{fig:ionosphere_map} shows the offset in position between the two catalogues for each source.
These offsets are also summarized in \fig~\ref{fig:ionosphere_offsets}.
They vary across the sky due to ionospheric structures which have been imaged and studied with the MWA \citep[e.g.][]{2015RaSc...50..574L}.
Among nighttime observations, this level of structure and magnitude are quite extreme, occurring only once in a 30-night study \citep{2017arXiv170704978J}.
However, over almost all of the field of view, the spatial variation in the offsets is sufficiently densely sampled to be fully captured.

We exploit this spatial correlation, using a Radial Basis Function (RBF) to characterize this vector field: determining the offset at any location to be the sum of all the offsets in the field, each weighted by the distance of each source from the point of interest to the inverse fourth power\footnote{For a homogeneous distribution of points weighting by distance$^{-2}$ will result in each concentric ring around the source being given equal weight. More negative powers weight the area around the source of interest more highly.}.
This makes it possible to map any apparent position to the true location consistent with the VLSSr catalogue.
To assess the effectiveness of this characterization, the offset at the location of each source was determined using an RBF which included all sources \emph{except} the one in question (see \fig~\ref{fig:ionosphere_map}).
This analysis suggests that the average offset of sources from their VLSSr location drops to an arcminute or less for almost all sources.
In reality, this is an overestimate in the errors of our corrections over most of the image since all of the sources with an error above the 95th percentile are either close to the edge of the image or are clearly discrepant from neighbouring sources, indicating a poor match or a difference in morphology between VLSSr and our observations, rather than genuine ionospheric structure.
Additionally many of our sources have counterparts in VLSSr, in which case the RBF will return the VLSSr location.

\subsubsection{Absolute Flux Density Calibration}
The closest band in the GLEAM survey \citet{2017MNRAS.464.1146H} was then used to set the flux density scale.
Unfortunately, approximately one quadrant of our field is not included in GLEAM, however coverage is sufficient to provide an accurate flux density scale.
The central frequencies agreed within 3\% so no attempt was made to correct for this.
The scaling was set by hand and the agreement after this has been applied was within errors.

\subsection{Time-series Imaging}
\label{sec:timeseries_imaging}
Next, each of the individual 0.5\,s time integrations which make up the 5-minute observation were imaged individually, again using WSClean.
Imaging parameters (pixels size, image dimensions) were kept the same as for the standard image (see \tab~\ref{tab:observations}) including imaging of both orthogonal polarizations and cleaning of each snapshot.
The image data were then re-ordered into a single file which was structured to make it efficient to extract either a single time series or a number of timeseries covering a small patch of sky (see \sect~\ref{app:image_cubes}).
Mean, variability, skew and kurtosis maps were then constructed for each frequency as described in \sect~\ref{sec:var_imaging}.

\section{Results}
\label{sec:results}
\begin{figure*}
  \includegraphics[width=\textwidth]{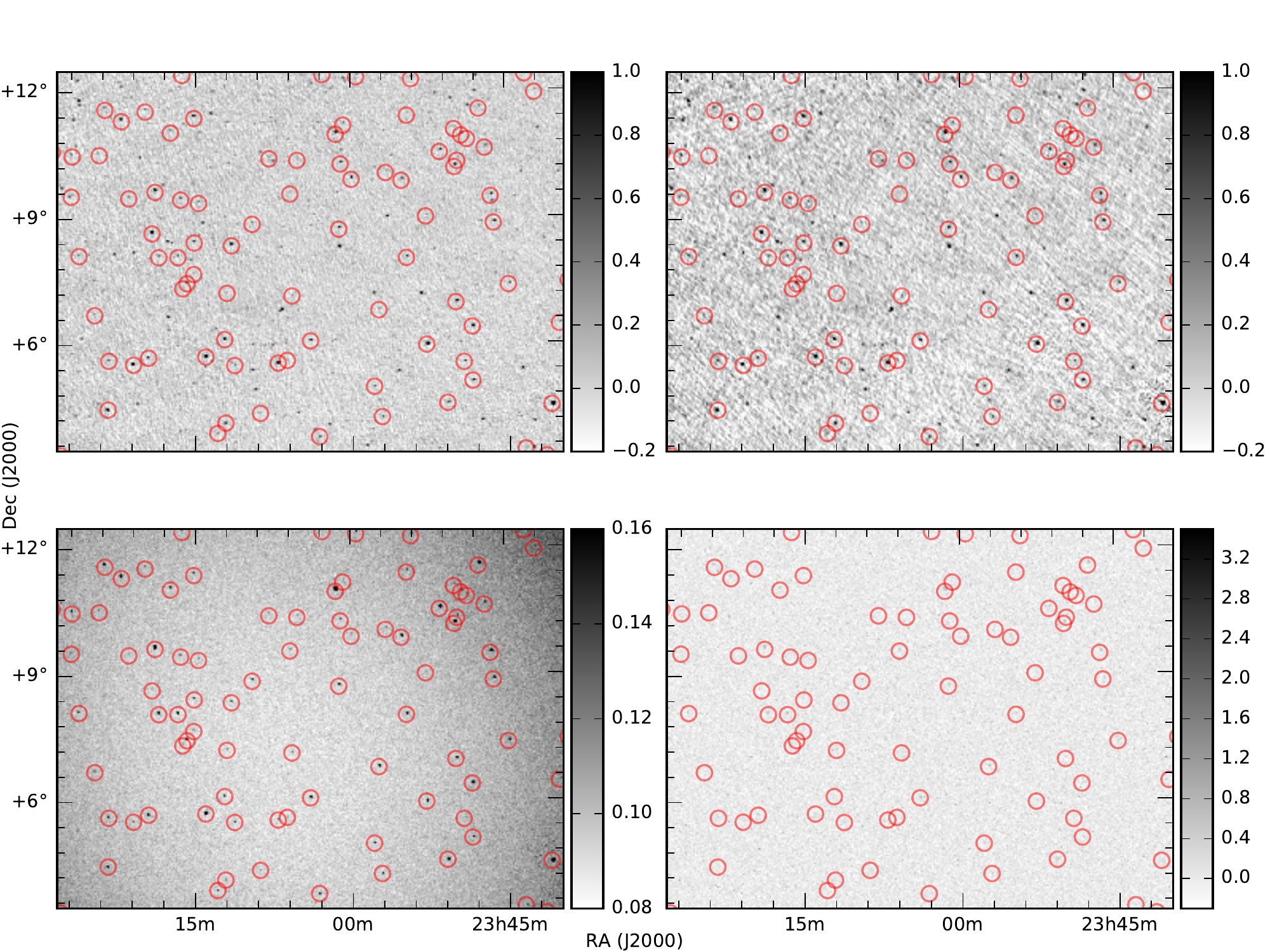}\\
  \caption{A representative area close to the centre of the high-band images: Top left: standard (Jy/beam), top right: mean (Jy/beam), bottom left: variability (Jy/beam), bottom right: skew. Catalogued scintillators indicated.}
  \label{fig:ips_image_hi}
\end{figure*}
\begin{figure*}
  \includegraphics[width=\textwidth]{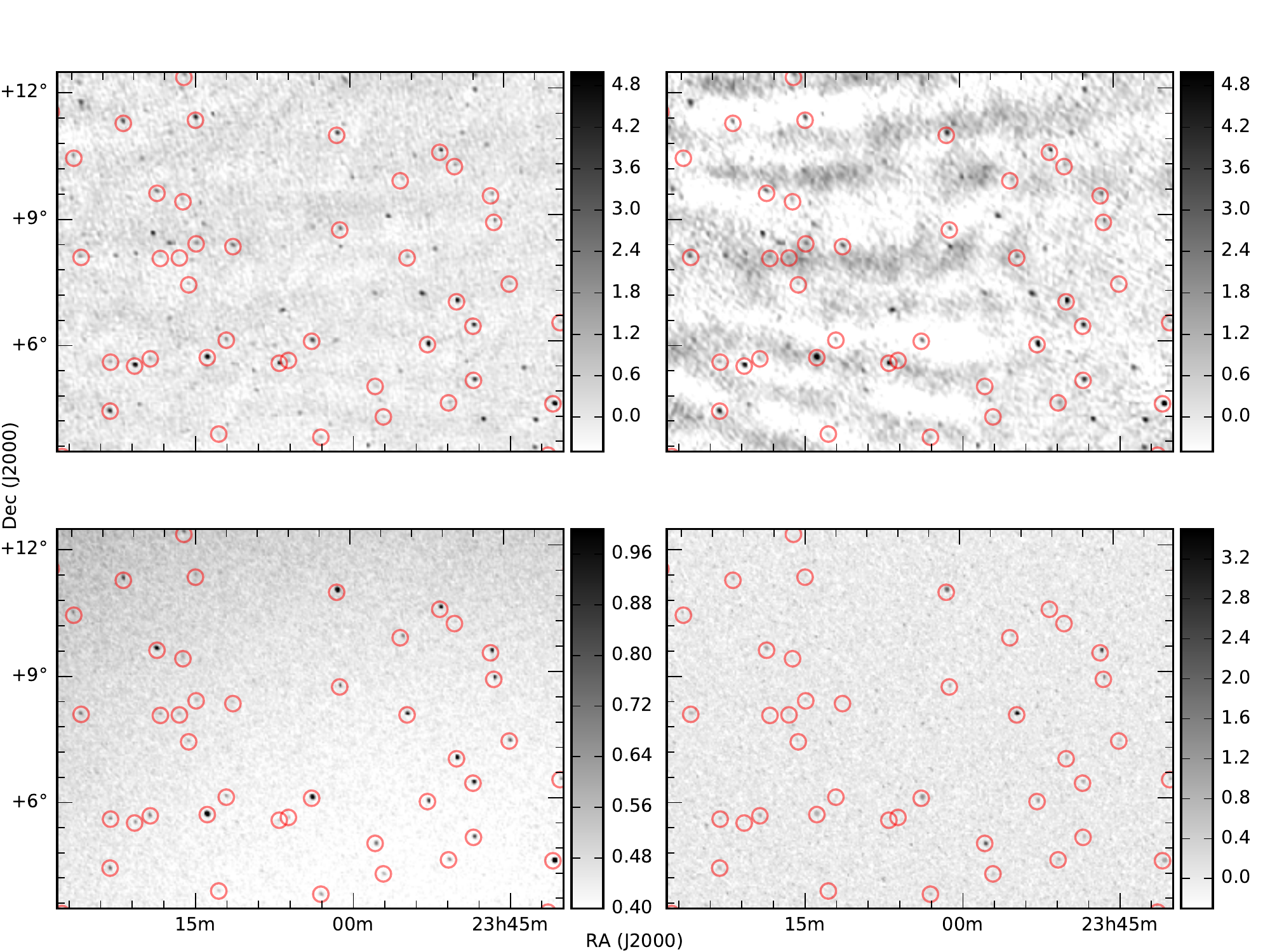}\\
  \caption{A representative area close to the centre of the low-band images: Top left: standard (Jy/beam), top right: mean (Jy/beam), bottom left: variability (Jy/beam), bottom right: skew. Catalogued scintillators indicated.}
  \label{fig:ips_image_lo}
\end{figure*}
\Figs~\ref{fig:ips_image_hi}\&\ref{fig:ips_image_lo} show a representative area close to the field centre as it appears in the standard, mean, variability and skew images.
The results are encouraging: the mean and standard image are similar, except that the standard image is far better deconvolved (this is to be expected since individual 0.5\,s images cannot be cleaned as deeply as an image formed from a 5-minute integration).
The variability image has the expected positive bias due to system noise: around 140\,mJy in the high band, and 525\,mJy in the low band.
The former is comparable with the 128\,mJy noise level at 150\,MHz predicted by \citet{2013PASA...30....7T} for our observing parameters (including 35\% sensitivity loss due to off-zentith pointing).
Except close to bright scintillators, where the rms is increased by sidelobe confusion, the ratio of the positive bias to the (spatial) rms is just a few percent less than the expected $\sqrt{560}$ over almost all of the map, suggesting that system noise dominates.
This implies 5$\sigma$ detection limits at the centres of the fields of 100\,mJy and 375mJy for the high and low bands respectively.

The variability image also contains what appear to be unresolved sources (i.e. their morphology matches the PSF).
Many of the brightest sources in the standard image are invisible in the variability image, however all of the sources in the variability image are apparent in the standard image, though some are $<5\sigma$.
The skew maps are much more noisy, however many sources visible in the variability image do appear to have clear counterparts (more so in the low-band).
These qualitative features are all consistent with the sources in the variability image being those showing IPS.

\subsection{Ionospheric effects in the image domain}
\label{sec:ionosphere_image}
\begin{figure*}
  \includegraphics[width=\textwidth]{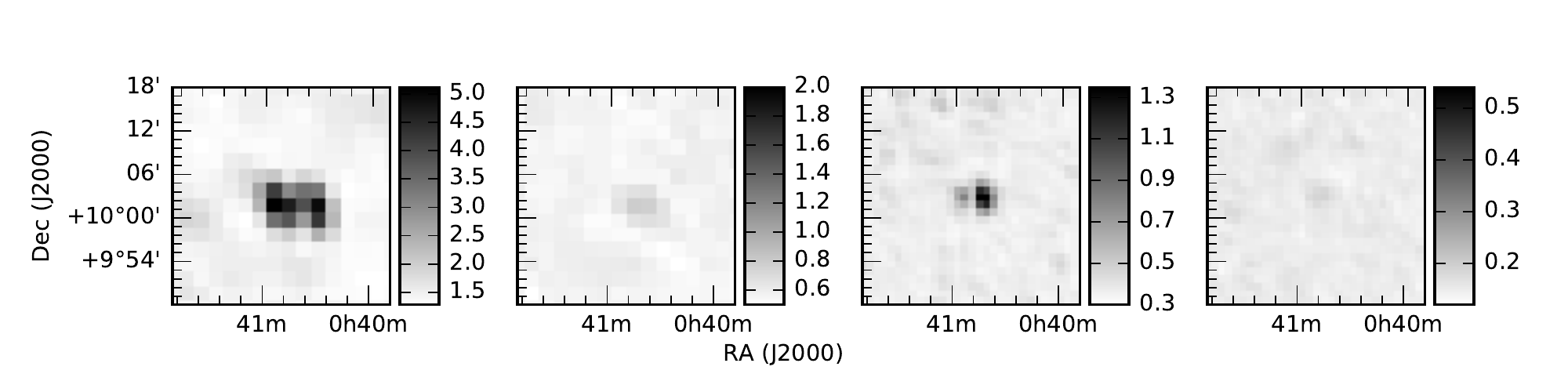} \\
  \caption{3C18 in variability images (all Jy/beam). Left to right: low band, no time-domain filtering; low band, normal time-domain filtering; high-band, no time-domain filtering; high-band normal time-domain filtering}
  \label{fig:ionosphere_image}
\end{figure*}
We also produced a variability image with no time-domain filtering in order to see the potential effects of the ionosphere on a variability image.
Since ionospheric scintillation manifests on spatial scales $\sim$10\arcmin, this scintillation will generally be resolved by the MWA, with ionospheric scintillation causing distortion in the image plane.
This is illustrated in \fig~\ref{fig:ionosphere_image}, which shows source 3C18 (the same source used in \fig~\ref{fig:non_ips_source}) as it appears in the standard image, and in variability images with various time-domain filtering applied.
With no filtering of ionospheric scintillation timescales, the source appears to break into two with an East and West component.
Inspection of images of the source made on a 10\,s imaging cadence reveal that the variability of the source is dominated by East-West movement of the source by a fraction of a beam.
The variance is therefore highest on the East and West flanks where the gradient of the PSF is steepest.

After filtering, a weak source is still visible in the variability image, however the extended morphology is now gone.
If this is indeed due to IPS from a compact component in 3C18, then this would be consistent with higher frequency radio observations, which show a compact core with lobes approximately 1\arcmin\ in extent \citep{1993MNRAS.263.1023M}.
The core is flat spectrum with a flux density $\sim$70\,mJy at 5\,GHz and above \citep[][and references therein]{2008ApJ...678..712D}.
This is somewhat less than the scintillating flux density measured.
It is possible that the ionosphere is still contributing slightly to the measured scintillating flux density, or that hotspots in the lobes, whose alignment to the core is close to perpendicular to the solar wind direction, are contributing to the IPS power.  

\subsection{Source finding}
\label{sec:sourcefinding}
Aegean was used to identify the sources (at 5-$\sigma$ and above) in the variability images.
It was necessary to remove a number of spurious detections near the edge of the map, (and the Sun in the case of the low band).
Some of these were artifacts on the very edge of the map; others were aliases from just beyond the field of view, most notably \object{3C48}, one of the brightest IPS sources in the sky, which lies just beyond the edge of the high-band image.

It was also necessary to filter out the sidelobes of scintillating sources.
To disambiguate these from other scintillating sources, we exploited the fact that the scintils of each source are independent.
A ``correlation image'' was produced for each of the brightest scintillators, where each pixel value in the image was the correlation coefficient of the time series of the source in question, and the time series corresponding to the pixel.
Sidelobes of scintillating sources matched in both position and morphology with features in the correlation images, whereas independent scintillating sources showed no signature at all in the correlation images.
This property of scintillating sources means that in a variability image, unlike a standard interferometry image, sidelobes of different sources cannot constructively interfere, and so the effects of sidelobe confusion are more restricted to regions very close to the brightest scintillating sources.

\subsection{Cross matching}
\label{sec:crossmatching}
Next, the source positions were corrected for ionospheric effects, and the high- and low-band source catalogues were cross matched with the TGSSADR catalogue \citep{2017A&A...598A..78I}.
Additionally, a crossmatch was made to a dummy catalogue where the RA had been shifted by 5\degr, to assess the probability of spurious crossmatches.
The cross matching was done using TOPCAT \citep{Taylor:2005}, and initially, all cases where the fitted ellipse of one source overlapped with the other were recorded.
These ellipses are considerably larger than the uncertainties in the centroid position, however, we rejected any detections which did not have a TGSS match within 1\arcmin (high-band) or 2\arcmin (low-band).
For the most part, these detections without counterparts were restricted to the very edge of the map where the density of sources in our standard image is low and the RBF model fails to resolve ionospheric structure (see \fig~\ref{fig:ionosphere_map}). 
In the high band, there are a handful of detections with no plausible counterpart, however all but one of these are $<6\sigma$.


The above procedure should leave us with a reliable catalogue of sources which are showing excessive variability on IPS timescales.
All that remains is to determine that this variability is indeed IPS.
The power spectrum of each source was examined for the spectral signature of IPS.
All but 10 sources had a power spectrum consistent with IPS.
These 10 sources, all from the low band image, appear to be associated with sources which are extended on arcminute scales (i.e. they are resolved by the MWA). 
On further inspection, there appears to be an excess of noise of approximately $(0.15\pm0.15)\%$ of the mean flux density of a source in the high band, and $(0.5\pm0.5)\%$ of the mean flux density in the low band. 
We subtract this excess from our scintillating flux density estimates and increase our errors to reflect the uncertaint on this excess noise.
However this only has a significant effect on a few sources with very low scintillation indices.

Finally, we consider those sources which had multiple matches (77 sources in the high-band catalogue and 71~in the low-band catalogue)
We compared the likelihood of all potential matches, assuming the astrometric errors in our catalogue follow a circular Gaussian distribution with $\sigma$ given by the major axis of the uncertainty ellipse shown in  \fig~\ref{fig:ionosphere_offsets}.
Just seven sources in the high-band catalogue have an plausible alternative match (p>0.05), along with 29~in the low-band catalogue, (including 3 sources with 3 plausible crossmatches).
We list these alternative crossmatches in \tabs~\ref{tab:cat_hi}\&\ref{tab:cat_lo}.

\begin{figure*}
  \includegraphics[width=\textwidth]{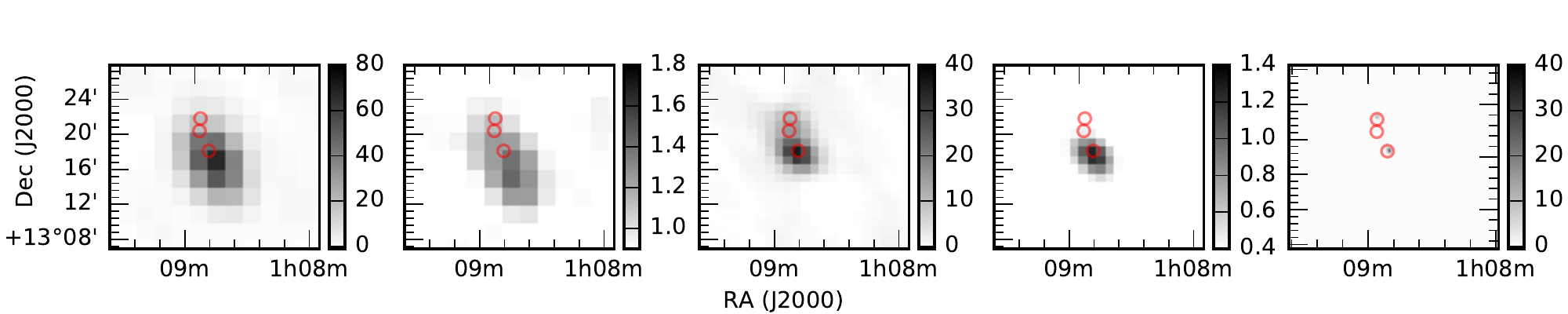} \\
  \caption{\object{3C33} (TGSSADR J010850.9+131840 in both catalogues) as it appears in the low-band standard image, low-band variability image, high-band standard image, high-band variability image, and in TGSS (all Jy/beam). Circles in each image indicate the location of TGSS sources. There is no ambiguity over which component is scintillating in either band.}
  \label{fig:3c33}
\end{figure*}
The fact that so few sources which have many potential matches are, in fact, ambiguous reflects the fact that after ionospheric corrections, the accuracy with which we can locate our scintillating sources far exceeds the native resolution of the array.
This is further illustrated in \fig~\ref{fig:3c33} which shows that the scintillation detection of 3C33 comes from the Southern lobe, most likely from a hotspot.

For the low-band catalogue, the best matches were then matched with the VLSSr catalogue.
Just one detection lacked a VLSSr counterpart.
The uncertainties on the VLSSr flux density were combined in quadrature with a fractional error of 10\%, as is done for TGSS, to reflect the difficulties of maintaining a uniform flux density scale across the sky in low-frequency surveys \citep[see][\sect~4.5]{2017A&A...598A..78I}.

\subsection{Calculation of scintillation indices}
\label{sec:scint_index}
We have shown that we can conduct a survey of all scintillating sources within our field.
However, in order to relate the magnitude of IPS that we see for each of our sources to physical parameters such as the solar wind density or the source size, we must determine the scintillation index of each source, i.e. the ratio of scintillating flux density to total flux density.

We note that the scintillating flux density as measured in the variability image is a somewhat biased estimate of the scintillating flux density.
The filters which we apply to our timeseries will filter out some of the IPS signal, and to some extent this will vary on a source by source basis, since the IPS timescale varies based on source size and solar wind speed.

There are three main choices for a flux density with which to normalize our measurement.
We could use either the mean of the timeseries, the value of the corresponding pixel in the standard image (the synthesis image made using the full 5 minutes of data), or a flux density value from VLSSr or TGSS.
Using the mean would eliminate errors due to the primary beam and imaging effects such as clean bias.
However, the errors in the flux densities derived from the mean image are extremely high due to sidelobe confusion.
The standard image has much lower sidelobe confusion, however there may be discrepancies due to the differing imaging processes, and there are still many scintillating sources without a strong detection.
Using flux densities from another catalogue will incur a potential error for any variable source (though such sources are exceedingly rare: see, e.g. \citet{2017MNRAS.466.1944M})
The flux densities from the matched catalogue do not match precisely in observing frequency (this would incur an error of 5\% for a source with a spectral index of magnitude 0.7), however these measurements have by far the lowest noise, and are the only reasonable choice for the many scintillating sources which are well below 5-$\sigma$ in the mean and standard images.
For those sources above 10-$\sigma$ in the mean and continuum image, their mean flux densities agree to within 10\%, except for a subset which are much weaker in VLSSr/TGSS, presumably due to resolution effects.

\Tabs~\ref{tab:cat_hi}\&\ref{tab:cat_lo} provide the measured scintillating flux density for each source, as well as the flux density from TGSS/VLSSr to facilitate the calculation of the scintillation index if required.
\subsection{Calculation of compact flux densities of sources}
\label{sec:compact flux}
The expected scintillation index $m$ for an unresolved source is approximately
\begin{equation}
	m=0.06\lambda^{1.0}p^{-1.6}
	\label{eqn:rickett}
\end{equation}
where $\lambda$ is the wavelength in m, and $p$ is the point of closest approach of the line of sight to the Sun in AU (the sine of the solar elongation of the source).

This relationship has been shown to hold over a wide range of observing frequencies and solar elongations and over the solar cycle \citep{1969P&SS...17..313H,1973JGR....78.1543R,1993SoPh..148..153M}, while the source remains in the weak scintillation regime (i.e. the expected scintillation index is less than unity).
However solar wind conditions along the line of sight can either reduce or increase the scintillation index, with higher values indicative of a compact source which is showing enhanced scintillation due to increased turbulence along the line of sight.
Nonetheless, this relationship can be used to convert the measurements of \emph{scintillating} flux density given in \Tabs~\ref{tab:cat_lo}\&\ref{tab:cat_hi} into estimates of \emph{compact} flux density of the source.

\begin{landscape}
  \begin{table}
    \footnotesize
    \caption{High-band catalogue of scintillating sources (only the first 10 are shown). Table columns are defined as follows:
    (1) Source name of best TGSSADR match;
    (2) Right Ascension J2000.0 of best match;
    (3) Declination J2000.0 of best match;
    (4) Probability that best match is correct match;
    (5) Source name of 2nd best match;
    (6) Probability that 2nd best match is correct match;
    (7) TGSSADR Flux Density of best match;
    (8) Error on (7);
    (9) Solar Elongation on date of observation;
   (10) Scintillating Flux Density measured;
   (11) Error on (10).
    }
    \centering
	\begin{tabular}{|l|r|r|r|c|r|r|r|r|r|r|}
	\hline
	  \multicolumn{1}{|c|}{name\_match1} &
	  \multicolumn{1}{c|}{RAJ2000\_match1} &
	  \multicolumn{1}{c|}{DEJ2000\_match1} &
	  \multicolumn{1}{c|}{p\_match1} &
	  \multicolumn{1}{c|}{name\_match2} &
	  \multicolumn{1}{c|}{p\_match2} &
	  \multicolumn{1}{c|}{S} &
	  \multicolumn{1}{c|}{e\_S} &
	  \multicolumn{1}{c|}{elongation} &
	  \multicolumn{1}{c|}{dS} &
	  \multicolumn{1}{c|}{e\_dS} \\
	  \multicolumn{1}{|c|}{} &
	  \multicolumn{1}{c|}{deg} &
	  \multicolumn{1}{c|}{deg} &
	  \multicolumn{1}{c|}{} &
	  \multicolumn{1}{c|}{} &
	  \multicolumn{1}{c|}{} &
	  \multicolumn{1}{c|}{Jy} &
	  \multicolumn{1}{c|}{Jy} &
	  \multicolumn{1}{c|}{deg} &
	  \multicolumn{1}{c|}{Jy} &
	  \multicolumn{1}{c|}{Jy} \\
	  \multicolumn{1}{|c|}{(1)} &
	  \multicolumn{1}{c|}{(2)} &
	  \multicolumn{1}{c|}{(3)} &
	  \multicolumn{1}{c|}{(4)} &
	  \multicolumn{1}{c|}{(5)} &
	  \multicolumn{1}{c|}{(6)} &
	  \multicolumn{1}{c|}{(7)} &
	  \multicolumn{1}{c|}{(8)} &
	  \multicolumn{1}{c|}{(9)} &
	  \multicolumn{1}{c|}{(10)} &
	  \multicolumn{1}{c|}{(11)} \\
	\hline
	  J000000.1--040243 & 0.0007 & --04.0454 & 1.000 & -- & 0.000 & 0.40 & 0.04 & 40 & 0.274 & 0.028 \\
	  J000002.9+095705 & 0.0121  & --09.9515 & 1.000 & -- & 0.000 & 1.57 & 0.16 & 35 & 0.105 & 0.014 \\
	  J000031.6--025143 & 0.1318 & --02.8622 & 1.000 & -- & 0.000 & 1.38 & 0.14 & 39 & 0.217 & 0.023 \\
	  J000047.1+111429 & 0.1963  & +11.2415  & 0.999 & -- & 0.000 & 0.90 & 0.09 & 35 & 0.120 & 0.014 \\
	  J000104.5+101928 & 0.2691  & +10.3246  & 1.000 & -- & 0.000 & 1.17 & 0.12 & 35 & 0.182 & 0.016 \\
	  J000115.5+084638 & 0.3148  & +08.7773  & 1.000 & -- & 0.000 & 2.37 & 0.24 & 35 & 0.398 & 0.030 \\
	  J000130.9+110140 & 0.3791  & +11.0280  & 1.000 & -- & 0.000 & 3.48 & 0.35 & 35 & 1.429 & 0.105 \\
	  J000132.7+145608 & 0.3864  & +14.9358  & 1.000 & -- & 0.000 & 1.36 & 0.14 & 34 & 0.249 & 0.021 \\
	  J000145.8+145806 & 0.4411  & +14.9684  & 1.000 & -- & 0.000 & 1.05 & 0.11 & 34 & 0.195 & 0.018 \\
	\hline\end{tabular}
    \label{tab:cat_hi}
  \end{table}
  \begin{table}
    \caption{Low-band catalogue of scintillating sources (only the first 10 are shown). Table columns are defined as follows:
    (1) Source name of best TGSSADR match;
    (2) Right Ascension J2000.0 of best match;
    (3) Declination J2000.0 of best match;
    (4) Probability that best match is correct match;
    (5) Source name of 2nd best match;
    (6) Probability that 2nd best match is correct match;
    (7) Source name of 3rd best match best match;
    (8) Probability that 3rd best match is correct match;
    (9) VLSSr Flux Density of best match;
   (10) Error on (9);
   (11) Solar Elongation on date of observation;
   (12) Scintillating Flux Density measured;
   (13) Error on (12).
    }
	\begin{tabular}{|l|r|r|r|c|r|c|r|r|r|r|r|r|}
	\hline
	  \multicolumn{1}{|c|}{name\_match1} &
	  \multicolumn{1}{c|}{RAJ2000\_match1} &
	  \multicolumn{1}{c|}{DEJ2000\_match1} &
	  \multicolumn{1}{c|}{p\_match1} &
	  \multicolumn{1}{c|}{name\_match2} &
	  \multicolumn{1}{c|}{p\_match2} &
	  \multicolumn{1}{c|}{name\_match3} &
	  \multicolumn{1}{c|}{p\_match3} &
	  \multicolumn{1}{c|}{S} &
	  \multicolumn{1}{c|}{e\_S} &
	  \multicolumn{1}{c|}{elongation} &
	  \multicolumn{1}{c|}{dS} &
	  \multicolumn{1}{c|}{e\_dS} \\
	  \multicolumn{1}{|c|}{} &
	  \multicolumn{1}{c|}{deg} &
	  \multicolumn{1}{c|}{deg} &
	  \multicolumn{1}{c|}{} &
	  \multicolumn{1}{c|}{} &
	  \multicolumn{1}{c|}{} &
	  \multicolumn{1}{c|}{} &
	  \multicolumn{1}{c|}{} &
	  \multicolumn{1}{c|}{Jy} &
	  \multicolumn{1}{c|}{Jy} &
	  \multicolumn{1}{c|}{deg} &
	  \multicolumn{1}{c|}{Jy} &
	  \multicolumn{1}{c|}{Jy} \\
	  \multicolumn{1}{|c|}{(1)} &
	  \multicolumn{1}{c|}{(2)} &
	  \multicolumn{1}{c|}{(3)} &
	  \multicolumn{1}{c|}{(4)} &
	  \multicolumn{1}{c|}{(5)} &
	  \multicolumn{1}{c|}{(6)} &
	  \multicolumn{1}{c|}{(7)} &
	  \multicolumn{1}{c|}{(8)} &
	  \multicolumn{1}{c|}{(9)} &
	  \multicolumn{1}{c|}{(10)} &
	  \multicolumn{1}{c|}{(11)} &
	  \multicolumn{1}{c|}{(12)} &
	  \multicolumn{1}{c|}{(13)} \\
	\hline
	  J000031.6--025143 & 0.1318 & --02.8622 & 0.999 & --               & 0.000 & -- & 0.000 &  3.12 & 0.33 & 39 & 0.86 & 0.08\\
	  J000040.0--142348 & 0.1671 & --14.3968 & 1.000 & --               & 0.000 & -- & 0.000 &  4.60 & 0.47 & 45 & 0.64 & 0.08\\
	  J000057.6--105431 & 0.2400 & --10.9089 & 0.999 & --               & 0.000 & -- & 0.000 & 10.91 & 1.09 & 43 & 2.95 & 0.26\\
	  J000105.4--165926 & 0.2726 & --16.9906 & 1.000 & --               & 0.000 & -- & 0.000 &  8.89 & 0.89 & 47 & 0.88 & 0.11\\
	  J000115.5+084638  & 0.3148 & +08.7773  & 1.000 & --               & 0.000 & -- & 0.000 &  4.40 & 0.45 & 35 & 1.05 & 0.10\\
	  J000123.1--132603 & 0.3467 & --13.4342 & 0.505 & J000126.0-132632 & 0.494 & -- & 0.000 &  1.94 & 0.21 & 45 & 0.85 & 0.08\\
	  J000130.9+110140  & 0.3791 & +11.0280  & 1.000 & --               & 0.000 & -- & 0.000 &  5.90 & 0.59 & 35 & 2.03 & 0.18\\
	  J000132.7+145608  & 0.3864 & +14.9358  & 0.997 & --               & 0.000 & -- & 0.000 &  1.83 & 0.20 & 34 & 0.70 & 0.08\\
	  J000221.7--140643 & 0.5906 & --14.1121 & 0.983 & --               & 0.000 & -- & 0.000 &  5.53 & 0.56 & 45 & 0.69 & 0.08\\
	\hline\end{tabular}
    \label{tab:cat_lo}
  \end{table}
\end{landscape}

\section{Discussion}
\label{sec:discussion}
\subsection{Scintillation indices}
\begin{figure*}
    \centering
        \includegraphics[scale=0.8]{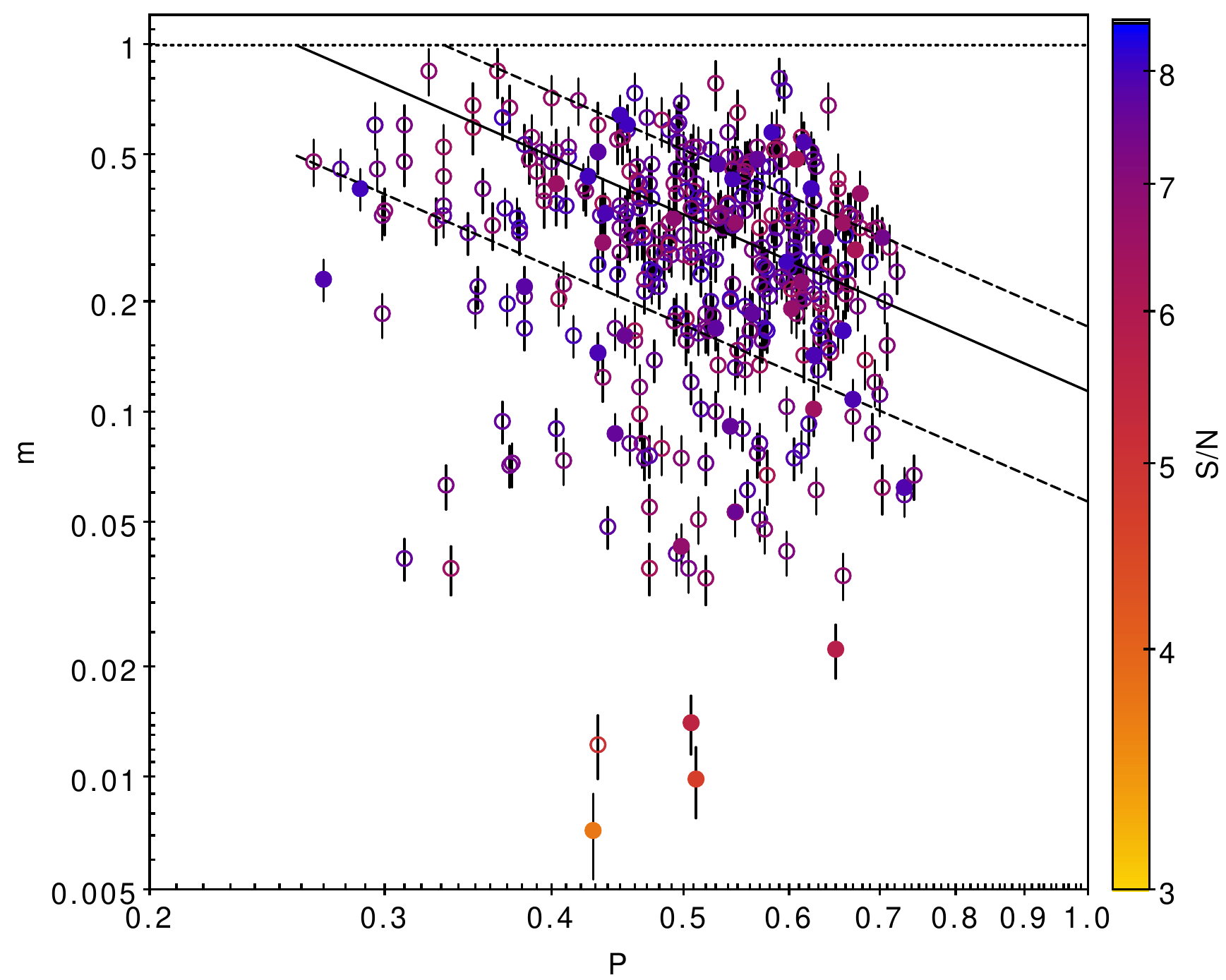}
        \includegraphics[scale=0.8]{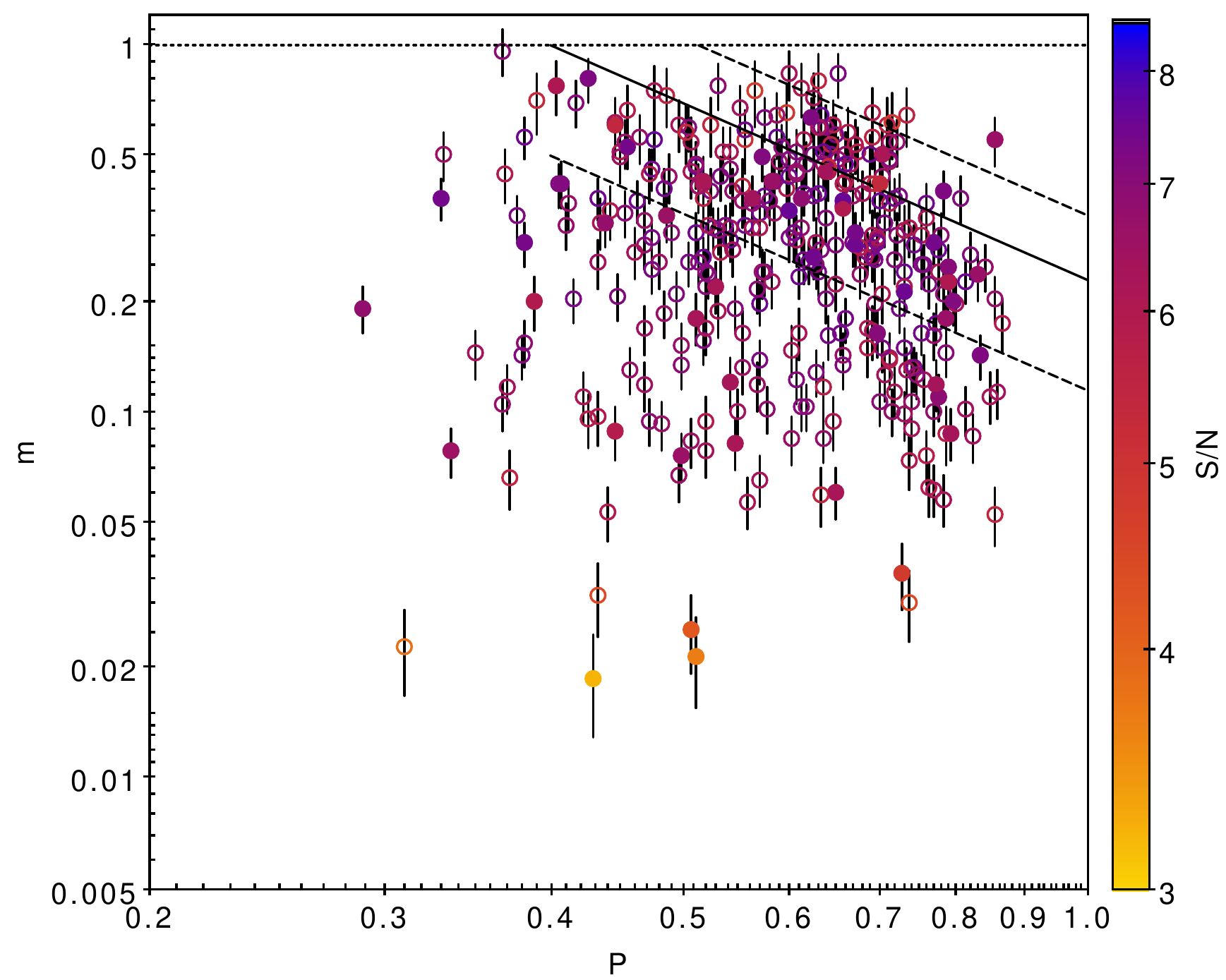}
    \caption{Scintillation indices ($m$) as a function of impact factor ($p$). Top: high band. Bottom: low band. Solid line shows the expected scintillation index for an unresolved source. Dotted lines  50\% and 150\% of the expected scintillation index. Filled circles indicate known VLBI sources.}
    \label{fig:scint_index}
\end{figure*}
\Fig~\ref{fig:scint_index} shows the scintillation indices for all sources as a function of impact factor.
Also shown is the expected scintillation index for an unresolved source, given by \eqn~\ref{eqn:rickett}.
The ratio of the scintillation index to that expected is known as the  g-level or $g$.
For an ideal point source, values of $g$ of 0.8--1.2 are typical in relation to large solar wind structures arising from active regions on the Sun \citep{1995GeoRL..22..643H}, and values of $g$ from 0.5--2.5 are within the range of observed values (\citealp{1986P&SS...34...93T}; see also \citealp{1987MNRAS.229..589P} \fig~4 for extensive daily data on a range of IPS sources).

For our sources, $g$ cannot be computed, since we cannot separate the effects of solar wind and source size from a single observation. 
However most of the observed scintillation indices lie between 50\% and 150\% of that expected for an unresolved source under average solar wind conditions.
This indicates that for most sources, the source's total flux density comes entirely or mostly from a compact component.
There is a small but significant number of sources which show a scintillation index $>$50\% higher than that expected.
Many of these sources lie at a common solar elongation in both bands, which is suggestive of space weather effects.
There is also a significant population of sources where only a small fraction of the source's total flux density is compact.
These could consist of sources with a highly compact component surrounded by more extended emission, or slightly extended sources with sizes $\sim$1\arcsec.
One of these sources is 3C18, discussed extensively in \sect~\ref{sec:ionosphere_image}.
\subsection{Comparison with data from other IPS observations}
\begin{figure*}
  \includegraphics[scale=0.45]{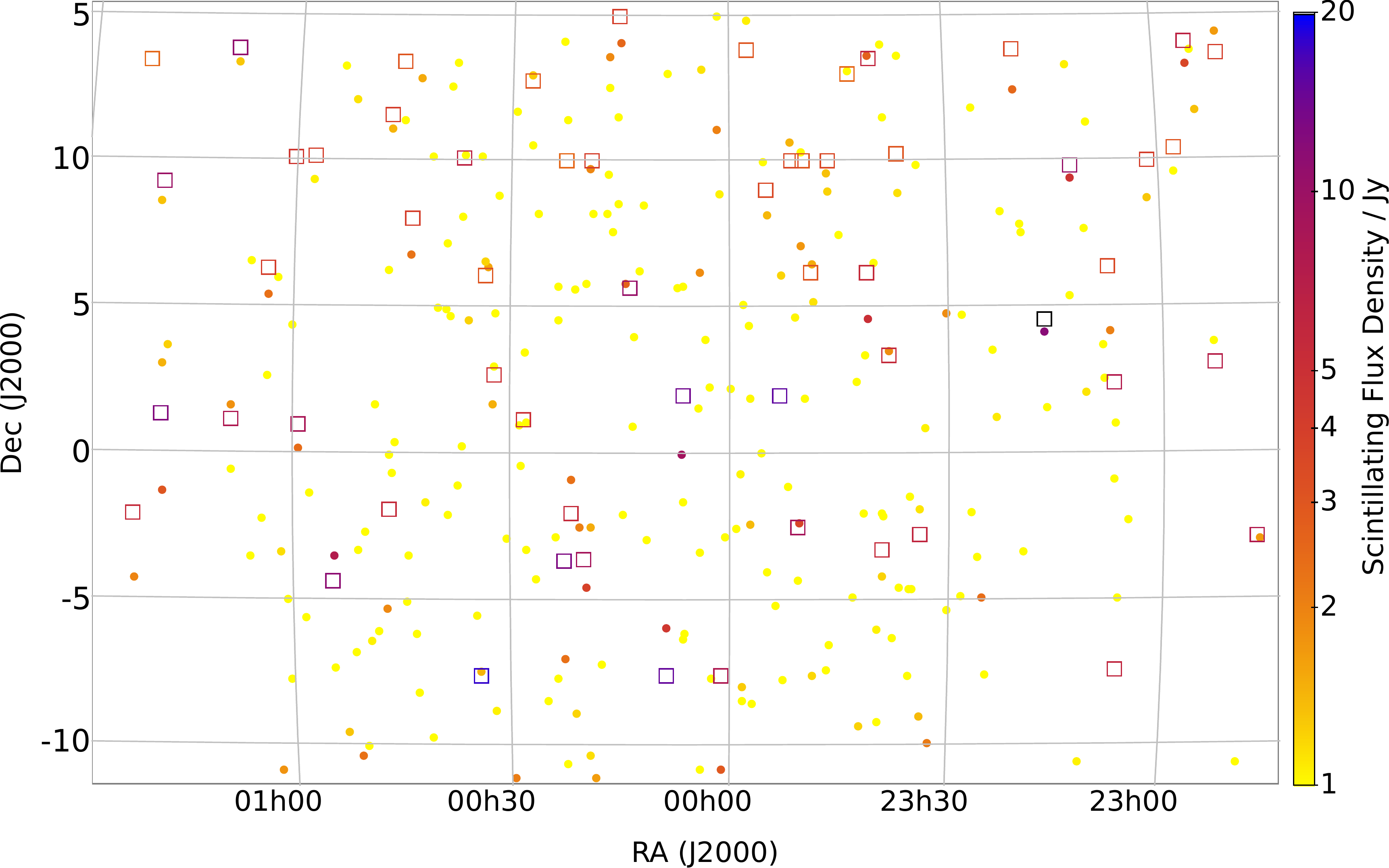}\\
  \caption{Comparison of MWA low-band catalogue (79\,MHz) with the Cambridge IPS survey at 81.5\,MHz for an overlapping region. Filled circles show MWA IPS sources, open squares show Cambridge sources. Colour scale is scintillating flux density (as modelled from observations at multiple solar elongations for Cambridge; as observed for MWA).}
  \label{fig:cambridge}
\end{figure*}
\Fig~\ref{fig:cambridge} compares our low-band catalogue with the Cambridge IPS survey \citep{1987MNRAS.229..589P}, conducted at a very similar observing frequency.
The agreement between the two catalogues is very good within the region plotted \emph{if} declination errors of a degree or more are allowed for (note that this region is close to the southern limit of the Cambridge survey, where the declination resolution of the instrument is at its poorest).
The agreement in the scintillating flux densities is reasonable considering that the Cambridge flux densities were determined from a fit to multiple observations at different solar elongations, whereas ours come from a single observation.

\begin{figure*}
  \includegraphics[width=0.7\textwidth]{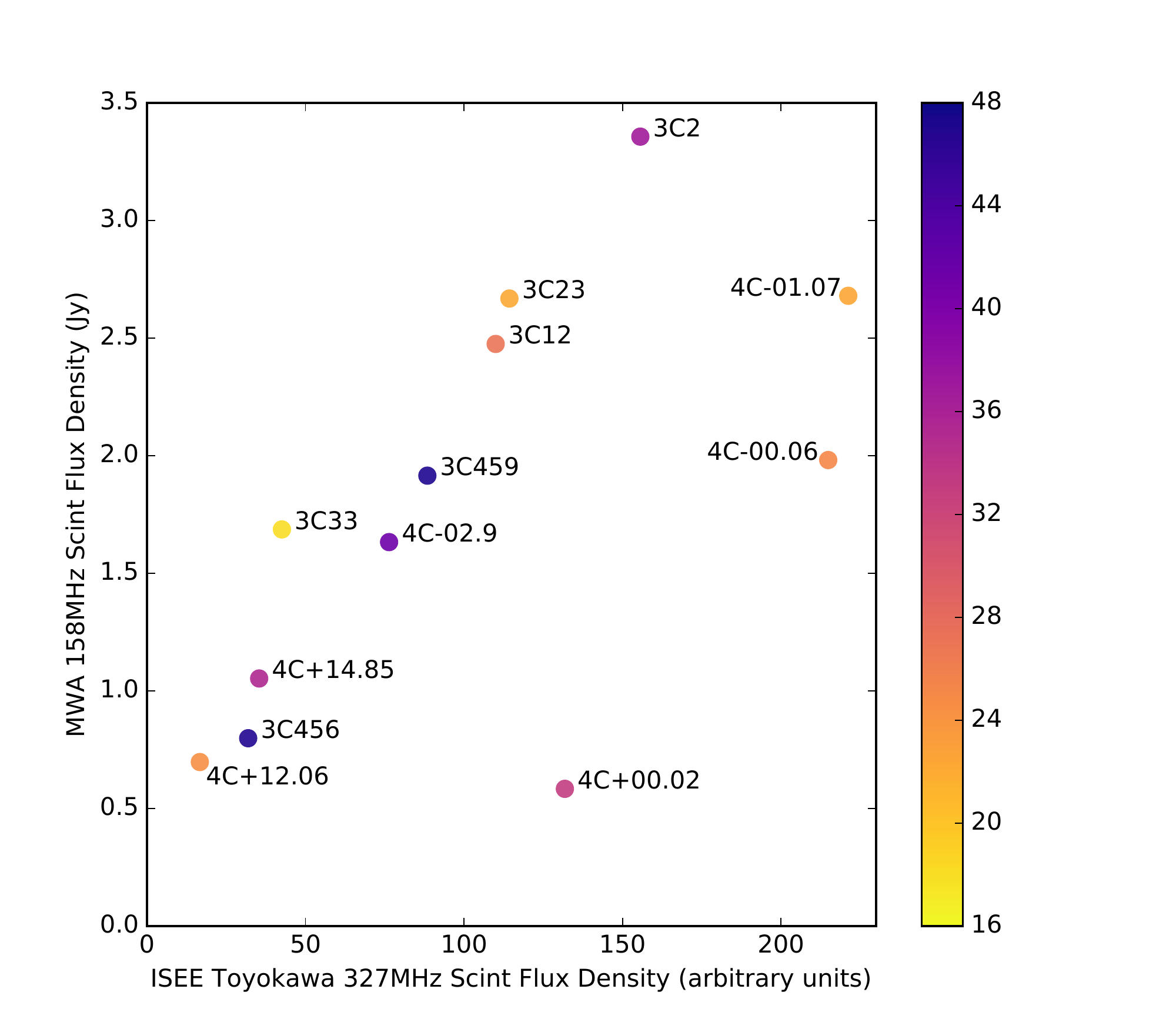} \\
  \caption{Comparison of MWA high-band scintillating flux density (Jy) with near-contemporaneous data from the Toyokawa station of the ISEE array.}
  \label{fig:isee}
\end{figure*}
\begin{figure*}
  \includegraphics[width=0.7\textwidth]{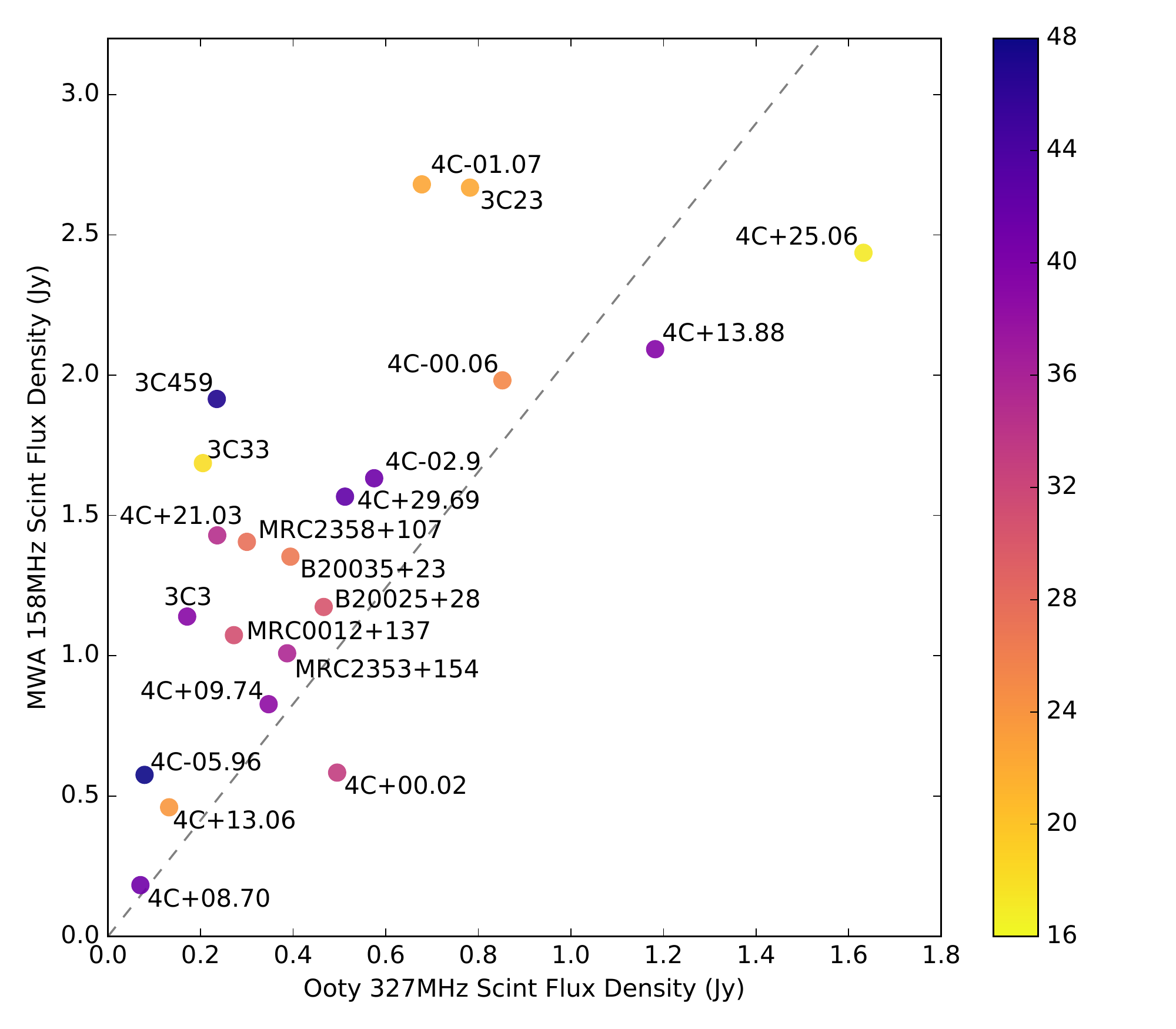} \\
  \caption{Comparison of MWA high-band scintillating flux density (Jy) with 327\,MHz measurements taken by Ooty approximately 8 hours later. Line shows expected gradient (see \eqn~\ref{eqn:rickett})}
  \label{fig:ooty}
\end{figure*}
%
\Fig~\ref{fig:isee} compares our high-band scintillating flux densities with 327\,MHz data from the Toyokawa station of the ISEE array (all taken within 80 minutes of our observations).
The scintillating power recorded by Toyokawa is essentially a signal-to-noise ratio which is expected to be proportional to the scintillating flux density.
In spite of the factor-of-two difference in observing frequency, which can be expected to change both the mean flux density and scintillation index of the sources considerably, there is a clear linear relationship between the measurements for the two instruments for all but three of the sources.

\Fig~\ref{fig:ooty} compares our high-band scintillating flux densities with 327\,MHz data taken by Ooty approximately 8 hours later.
These data shows a similar pattern to the ISEE data, however now we are comparing absolute flux densities we can see that the expected relationship given by \eqn~\ref{eqn:rickett} is reasonable.
If anything, this relationship underestimates the MWA scintillating flux density.
This is to be expected, since although compact sources might be expected to be flat-spectrum, astrophysical sources which are brighter at higher frequencies are extremely rare.
Additionally, the most extreme outlier, which is much brighter at the higher frequency, is also the closest source to the Sun, where it will be in the strong scattering regime at the higher frequency.

\subsection{Implications for Astrophysical studies}
Sources which are compact on arcsecond scales or smaller are interesting for a number of reasons.
Compact sources are necessary in order to calibrate arrays with long baselines.
Only compact sources are suitable for absorption studies (no significant absorption will be seen if the source is large relative to the absorbing structure).
Linear size is also an important astrophysical quantity in understanding many classes of radio source \citep[e.g.][use it to understand the evolution of radio galaxies]{1997AJ....113..148O}.
Additionally, Only IPS sources are likely to show variability, either intrinsic (an object 1 light-day across at 1\,kPc would be $<$1\arcsec) or due to interstellar scintillation (since this requires even smaller sources than IPS).
Any transient source would also show IPS (if at an appropriate solar elongation) for the same reason.

Our results also show that high-resolution studies at low frequencies do not simply rediscover the compact objects already identified via VLBI studies at higher frequencies. 
Only 44/377 high-band scintillators and 53/353 low-band scintillators appear in the ``Radio Fundamental Catalogue'' of all radio sources detected in Geodetic VLBI campaigns \footnote{version 2015c. See \url{http://astrogeo.org/rfc/}} (see \Fig~\ref{fig:scint_index}).
Unfortunately, although VLBI at GHz frequencies is routine, observing at lower frequencies is far more challenging, since the ionosphere changes significantly from antenna to antenna, as well as from source to source. 
Furthermore, only targeted searches are typically made, due to the impracticality of imaging even modestly large fields with arcsecond resolution.
IPS sidesteps both these problems, since it can be achieved using an array whose dimensions fit within a single isoplanatic patch of the ionosphere, and requires only arcminute-resolution synthesis imaging.

Nonetheless, targeted VLBI surveys have been shown to be feasible at metric wavelengths.
\citet{2015A&A...574A..73M} surveyed 630 carefully selected sources in two hours with International LOFAR at 140\,MHz.
From their detection rate, they estimated that there is one suitable primary calibrator per square degree of sky.
Although their observations are not calibrated in flux density, they estimate that a $\sim$100\,mJy compact source would be sufficiently bright to act as a primary calibrator.
\citet{2016A&A...595A..86J} refine this to 1 source per square degree for the short international baselines ($\sim$1\arcsec\ resolution) and just 0.5 sources per square degree for the longer international baselines ($\sim$0.5\arcsec\ resolution).
At the centre of our high-band field we reach a sky density of approximately 1 source per square degree, with the weakest sources detected likely to have compact flux densities $\sim$330\,mJy (using \eqn~\ref{eqn:rickett} to estimate compact flux density from scintillating flux density).
Some caution should be taken in interpreting this result, since variations in the solar wind will introduce a positive bias into our estimates of compact flux density for sources close to the detection limit \citep{1913MNRAS..73..359E}.
Furthermore, IPS sources may not necessarily be suitable calibrators if they have complicated extended structure.
Nonetheless, our observations suggest that the International LOFAR observations may underestimate either the compact source counts, or the calibrator flux density required to calibrate long baselines.
One reason may be that the preselection of particular sources has meant that many potential calibrators have been missed. 
This underlines the value of the IPS imaging technique in detecting and localize \emph{all} scintillating sources within the field of view (down to a well-defined flux density limit).
A more detailed analysis of the source counts of compact sources revealed by our IPS observations is underway \citep{2017bChhetri}.

\begin{figure}
  \includegraphics[scale=0.6]{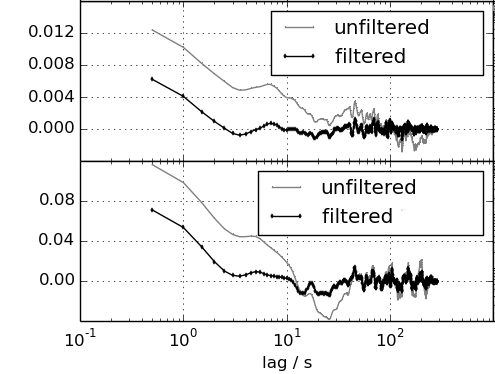}
  \caption{Autocorrelation Functions for \object{3C459}. Top: high band. Bottom: low band.}
  \label{fig:double}
\end{figure}
The large fraction of sources which show structure on 0.5\arcsec--1\arcsec\ scales noted by \citet{2016A&A...595A..86J} is consistent with our finding that a large number of sources have a smaller scintillation index than would be expected for a compact source.
This could be confirmed by an analysis of the power spectrum of the IPS, since such large source extent would be equivalent to convolving our time series with a Gaussian several samples wide.
The IPS signal encodes more detail on the source brightness distribution that simply the source's extent, however.
3C459, one of the brightest sources in our field, is a double source, with the components approximate 8\arcsec apart on a roughly E-W orientation \citep[see e.g.][]{1996ApJS..107..175R}. 
The autocorrelation function of the time series at both frequencies shows a clear bump at around 7 seconds lag (see \fig~\ref{fig:double}).
This can be explained by the solar wind flowing across one source component and then the other.
Accounting for the orientation of the components with respect to the solar wind velocity vector, this would imply a solar wind velocity of approximately 500km\,s$^{-1}$, which is reasonable. 
Using multiple observations at different solar elongations, the orientation of the double could be determined for sources off the ecliptic.
Once the double morphology is parameterized, the location of the bump in the ACF provides an estimate of the solar wind speed.

Together, the MWA (via IPS studies) and LOFAR (via the international baselines) can provide information on the sub-arcsecond sky at low frequencies across the entire celestial sphere with broadly similar resolution for each.
IPS studies are limited to sources within around 45\degr\ of the ecliptic\footnote{For in-depth studies across a range of frequencies, however some information could be gleaned for sources even 90\degr\ of the ecliptic: i.e. the full sky.}, however this covers a large swath of the Southern sky, including the Galactic centre and the South Galactic Cap.
In the SKA era, IPS studies in the Southern Hemisphere will have two important roles: firstly they will provide an accurate sky model of compact sources, important for calibration of long baselines.
This can be achieved without the longest baselines, vastly simplifying ionospheric calibration, so this work could be done by precursors (such as an upgraded MWA) or by a partially commissioned SKA-low.
Secondly IPS provides information on the structure of sources on spatial scales that lie between those probed by SKA-low baselines, and baselines \textit{between} the SKA-low and other facilities, such as the GMRT.

\begin{figure*}
  \includegraphics[width=\textwidth]{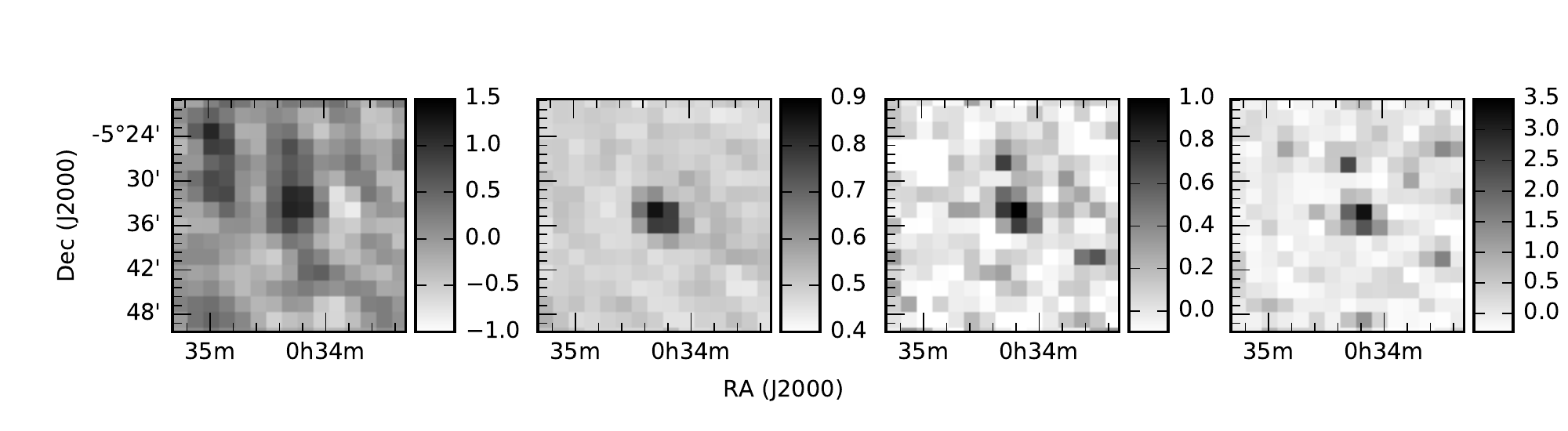} \\
  \caption{Pulsar \object{J0034-0534} (TGSSADR J003421.8-053437) as it appears in the low-band standard (Jy/beam), variability (Jy/beam), skew, and kurtosis images. The source is only approximately 3.8-$\sigma$ in the standard image, but is 17-$\sigma$ in the variability image.}
  \label{fig:pulsar}
\end{figure*}
Variability images have fundamentally different noise properties to the continuum images typically produced by radio interferometers.
At the centre of our field in our high band, we can detect a scintillating flux density of approx 100\,mJy at 5-$\sigma$ (vs 375\,mJy in the low band).
These numbers compare favourably with $5\times$ the RMS in the closest GLEAM bands (which have half the bandwidth, and perhaps double the integration time): 130mJy and 475mJy for 76\,MHz and 158\,MHz respectively.
This difference is even more stark when we compare with our own standard image, as illustrated in \fig~\ref{fig:pulsar}, which shows the only known pulsar that appears in either the high- or low-band catalogue.
The source is barely visible in our standard image, yet has a very strong detection in the variability image.
In total, there are 90 IPS sources in the high band image (and 99 in the low band image) which are $<5\sigma$ in the standard image.

The reason for this difference is that even snapshot images taken with the MWA are dominated by confusion noise.
In this case that includes very strong sidelobe confusion from the Sun.
In contrast, the variability images that we have produced are much further from the confusion limit, since compact emission makes up only a very small fraction of the total emission detected by the MWA (i.e. the vast majority of sources have a scintillation index $\ll$1) and any resolved emission (e.g. from the quiet Sun) is completely invisible.

Therefore IPS information assists in blind transient and variability studies in two different ways: first, with a catalogue of IPS sources covering the survey area, we can exclude the majority of sources from our analysis, since their linear size excludes them from being variable.
Secondly, we have a method which allows us to image the sky while filtering out any emission which does not come from sub-arcsecond scales, allowing us to probe below the confusion limit.

Unfortunately, while the current methodology is very effective at identifying hundreds of scintillating sources in our data, we have shown (in \sect~\ref{sec:scint_flux}) that there is little sensitivity to be gained from longer integrations, and the detection limit will be comparable with the sensitivity achievable in a single scintillation timescale.
Stacking analyses would suffer from the same $N^{1/4}$ limitation, though these may still allow detections well below the current sensitivity for large enough $N$.
Incremental gains may be found through improved calibration techniques such as those developed by \citet{2016MNRAS.458.1057O} and \citet{2017MNRAS.464.1146H}; adjusting the observation strategy to maximize instantaneous bandwidth and minimize instrumental noise; and by making the time-domain filter adaptive to the prevailing heliospheric conditions in each area of the sky. 
Additionally, the current statistics used for detection make no direct use of the exponential distribution of strong scintillation as a discriminant against Gaussian noise, nor are correlations in the IPS between frequency bands exploited.
Future investigations will focus on developing a detection metric which makes full use of this information to determine whether known sources are showing IPS, and to detect entirely new sources via their IPS signature alone.
For the former, prior knowledge of the locations of sources (whether they be detected in low-frequency radio surveys or at other wavelengths) could be used.

Future instruments or upgrades which increase the collecting area would of course allow deeper observations, as would increased bandwidth, at least where the scintillation bandwidth is wide, i.e. in the weak scintillation regime \citep[][]{2013SoPh..285..127F}.
As noted above, longer baselines than those used in this study would \emph{not} be required since confusion noise is not an issue and the current resolution is sufficient to allow cross matching to existing continuum catalogues.

Over most of the first half of 2016, we conducted daily IPS observations, with 10-minute observations of overlapping fields surrounding the Sun, while keeping the Sun itself in the null.
Reduction of these data is ongoing, however with multiple daily observations we can start to average over stochastic variations in scintillation index due to fluctuations in the solar wind, and by observing sources at multiple distances from the Sun we can start to model source sizes more accurately, producing a catalogue of \emph{compact} sources with a similar flux density limit to GLEAM.
Dual band observations will allow the spectral index of the compact component to be determined, as well as any change in size with frequency.

\subsection{Implications for Heliospheric science}
As noted in the previous section, we are able to detect one scintillating source per square degree, comparable to the deepest IPS surveys to-date \citep{2003A&A...403..555A}, with just 5 minutes of observation time.
This far exceeds the source density typically used for space weather studies.
Since our technique surveys wide fields for all scintillating sources, rather than targeting known sources, it is well-suited to conducting an all-sky survey of IPS sources.
Such a survey would be of great utility to existing facilities if they were upgraded to allow observations of multiple sources at once.

The MWA itself can also be used to generate IPS information for the Space Weather community.
Where IPS is used for space weather modelling, a number of observables are used to measure key physical parameters.
The first is the scintillation level of each source relative to its expected scintillation level ($g$).
The expected scintillation index varies from source to source (due to source structure) and depends on distance from the Sun.
Departures from the expected value (non-unity $g$) indicate unusually high or low turbulence ($\Delta N_e$), and $\Delta N_e$ is generally thought to be very closely related to $N_e$ \citep{1986P&SS...34...93T}.
We already have strong evidence for enhanced scintillation for some sources, however in order to calculate $g$ for all of the sources in the field, source structure needs to be accounted for.
Since source structure will be constant, all that needs to do be done in order to calculate $g$ is to observe the field a sufficient number of times to determine a median scintillation index for each source.
If this is done over a long enough interval that the solar elongation changes significantly, this would have to be taken into account.
The movement of areas of enhanced density or rarefaction (e.g. CMEs) can then be determined with extremely high resolution.
A feature moving at typical solar wind velocities will move across the sky at $\sim$1\degr\,hr$^{-1}$.
The IPS survey described in the previous section observed most sources more than once per day, but the MWA was only used with a duty cycle of approximately 0.25.
If the MWA were dedicated to IPS observations for a full day, we can imagine 6$\times$10\,minute observations being made every hour fully encircling the Sun with fields approximately 40\degr\ in diameter, probing solar elongations from 20--60\degr (except for when the Sun is within 45\degr\ of the horizon, when only the side of the Sun closest to the Zenith could be observed).
This is approach is technically feasible, although at the current time, calibration and imaging cannot be fully automated, and the computing facilities required to support regular observations of this magnitude would be considerable (see Appendix~\ref{app:computation}). 

In addition to $g$, velocity measurements are also important for tomographic reconstructions of the solar wind \citep[and references therein]{1998JGR...10312049J,2013SoPh..285..151J}.
This is often done using multi-station IPS, where the same scintils are tracked by two stations non-equal distances from the Sun, and the velocity can be determined from the delay \citep{1967Natur.213..343D,1971A&A....10..306L}.
However, velocities have also been measured using single station IPS using dual band measurements \citep{1983A&A...123..191S} or even using just a single station and single band \citep[e.g.][]{1990MNRAS.244..691M}, though there are a number of other parameters that have to be fitted for (as a minimum: source size, spectral index of turbulence and axial ratio of turbulence) and high signal to noise is required for an unambiguous result.
Prospects are therefore good for reliably determining the velocity using the dual-band technique, or using some combination of the multi-band capability and power spectrum fitting techniques, with results validated with near-contemporaneous observations from the ISEE array and/or the Ooty array for a subset of sources.

\subsection{Conclusions}

\begin{enumerate}
	\item We have developed a novel technique for measuring the IPS of many hundreds of sources simultaneously using high time resolution interferometric imaging.
	\item This technique can be used to conduct a full census of compact emission across the full field of view of the instrument.
	\item For space weather applications this means a much denser sampling of the heliosphere.
	\item For astrophysics, this provides a valuable complement to continuum surveys at low frequency, such as GLEAM and TGSS.
	\item For future instruments, such as SKA-low, this allows us to estimate the density of potential calibration sources.
	\item 98\% of sources in our 158\,MHz catalogue have an unambiguous counterpart in the TGSS ADR catalogue. Thus their location on the sky is known to 2\arcsec, enabling optical followup.
	\item When searching for a source by its variance alone, the detection threshold only increases by the fourth root of observing time. The same is true of bandwidth only if it exceeds the scintillation bandwidth. Otherwise sensitivity is proportional to the square root of bandwidth and number of baselines in the usual way.
\end{enumerate}
The next publication in this series \citep{2017Chhetri} will explore the astrophysical properties of these low-frequency compact sources in more detail.

\appendix
\section{Computational considerations}
\label{app:computation}
While there are numerous advantages to the synthesis imaging approach to IPS observations, especially for large-N arrays, the computational requirements and data products are large.
In this appendix we describe how to produce, store and access these data as efficiently as possible, provide estimates on the computational resources required, and indicate how future work might further improve efficiency. 

\subsection{Computationally efficient high time-resolution widefield synthesis imaging}
The exceptionally wide field of view of the MWA means that imaging accurately and efficiently is a complex problem.
Among the effects which make widefield imaging difficult are time-average and bandwidth smearing \citep{1999ASPC..180..371B}, w-term errors \citep{1999ASPC..180..383P}, and a changing phase calibration across the array due to the ionosphere. 
These effects can limit conventional imaging sensitivity, as any blurring of sources or residual sidelobes will increase confusion.
This has motivated the development of a number of tools \citep[e.g.][]{2008ISTSP...2..707M,2012ApJ...759...17S,2014MNRAS.444..606O} which address these issues.

These problems, however, are of little relevance to IPS imaging, at least with the field of view and baseline lengths used here.
Time average smearing is not an issue when imaging on very short timescales, neither are w-term errors for an instantaneously co-planar array (which the MWA is to a good approximation). 
Bandwidth smearing introduces decorrelation which ``smears'' sources in a radial direction (with the effect being proportional to the square of the distance from the phase centre), reducing their surface brightness though preserving their integrated flux density.
This change in the shape of the PSF can also cause deconvolution errors.

As discussed in \sect~\ref{sec:ionosphere_image}, ionospheric effects largely manifest themselves as shifts in the positions of sources.
The sidelobes of scintillating sources also change very noticeably across the field of view, almost certainly due to phase errors introduced by the ionosphere.
However, these do not have a significant negative effect, apart from reducing our sensitivity in the vicinity of bright sources, and introducing spurious detections, which are easily excluded (see \sect~\ref{sec:sourcefinding}).
This means that shortcuts can be taken to improve computational efficiency at the expense of imaging fidelity, for example averaging in frequency to reduce the number of visibilities, and reducing the number of facets/w-layers used to mitigate w-term errors.

We have found WSClean to be fast, accurate, convenient, and flexible enough to be easily adapted to our imaging problem.
The data discussed in this paper were reduced on a standard desktop machine over several days.
However, \tab~\ref{tab:computation} shows the time taken to execute the main data reduction tasks on a single node of the Pawsey Centre\footnote{\url{www.pawsey.org.au/our-systems/galaxy-technical-specifications/}} ``Galaxy'' computer (20$\times$3.0GHz CPU cores, 64\,GB memory).
\begin{table*}
  \footnotesize
  \centering
  \caption{\label{tab:computation}Time taken to execute various data reduction tasks on a single node of the Pawsey Centre ``Galaxy'' supercomputer, averaged across 12 observations. These observations have identical duration and correlator setup to the observations presented here (see \tab~\ref{tab:observations}), a similar dual-band setup has been used, and the zenith-angle is comparable. Note that obsdownload and cotter act on the full observation, whereas applysolution and wsclean have to be run separately on each of the bands.}
  \begin{tabular}{lrl}
    \hline
    Task           & Time (hours)  & Notes                          \\
    \hline                                                                                       
    obsdownload.py & 0.38       & Download data from MWA archive (executed on ``Zeus'', not Galaxy) \\
    cotter         & 0.61       & See \sect~\ref{sec:preprocessing}. In this case channels were averaged to 160\,kHz \\
    applysolutions & 0.60       & Apply calibration solution \\
    wsclean        & 5.65       & Imaging parameters identical to high band (\tab~\ref{tab:observations}) approx. 1000 CLEAN iterations \\
    \textbf{Total} & \textbf{7.76} & Currently, all of these tasks must be executed in serial. \\
    \hline
  \end{tabular}
\end{table*}
Data reduction time is currently dominated by imaging, which is in turn dominated by Fourier inversion of the visibilities (95\%, vs 5\% deconvolution).
WSClean already makes efficient use of multiple CPU cores, although if timesteps rather than w-layers could be imaged in parallel, this might lead to a speed up of an order of magnitude at the expense of poorer image fidelity. 
In the future GPUs might be used to do the imaging step even more rapidly.
Further gains could be found by consolidating the first three tasks.

The current minimum turnaround time $\sim$ 8 hours does not preclude useful space weather forecasting, since CME arrival times are usually measured in days rather than hours.
However this does assume that a calibration solution can be produced in an automated fashion relatively rapidly.
This has always been the goal \citep{2008ISTSP...2..707M} and will probably be a necessity for the next generation of radio interferometers.

\subsection{A Convenient Format for Image Cubes}
\label{app:image_cubes}
\Sect~\ref{sec:timeseries_imaging} describes the result of the imaging process: a 2D FITS image\footnote{\citep{1981A&AS...44..363W}} (of dimensions $N_x$=2400, $N_y$=2400), for each orthogonal polarization ($N_{p}$=2), for each timestep ($N_{t}$=560), for each of two bands.
For all images for a given band, the World Coordinate System \citep[WCS:][]{2002A&A...395.1061G} is identical.
In other words, a single pixel ($x$, $y$) maps to the same point on the celestial sphere for all images (setting aside ionospheric effects which are not accounted for).

Since $\sim$1000 images would have to be read in order to produce a single timeseries, some reorganization is required.
One option would be to reorganize these images into a single FITS file with dimensions [$N_y$, $N_x$, $N_{pol}$, $N_t$] (where we use the C-convention with the fastest-varying axis \emph{last}).
This would ensure that both polarizations of a time-series of a given pixel are stored contiguously, allowing rapid access.
The disadvantage of this approach is that we are often interested in accessing the time-series for near spatial neighbours (for example to examine off-source pixels).

The HDF5 format\footnote{\url{http://www.hdfgroup.org/HDF5}} deals with this problem by storing data in ``chunks'': a subset of an array where the size of each dimension is a factor of the size of the total array.
We store our data in an array of dimensions [$N_{pol}$, $N_y$, $N_x$, $N_{chan}$, $N_t$], where $N_{chan}$=1 and provides for any future dataset where multiple spectral channels are produced.
We then use chunks of dimensions [$N_{pol}$, $C$, $C$, $N_{chan}$, $N_t$] with $C$=24 for the high band and $C$=16 for the low band.
This means that in order to produce a sub-cube of spatial dimensions $C\times C$ a maximum of 4 chunks have to be read.

A single HDF5 file can store multiple multi-dimensional arrays (``datasets'') in a hierarchical containers (``groups''), so a single dataset can store data from multiple bands, along with other multi-dimensional data, such as primary beam images.
Metadata can also be attached each dataset as a set of key-value pairs (``attributes'').
The FITS header is stored in this way, and an astropy \citep{2013A&A...558A..33A} WCS object can be generated from it directly.
This means that summoning a timeseries can be made as simple as providing a right ascension and declination, with primary-beam corrections being made on-the-fly if required.

This data structure makes it extremely convenient to access the data, however the data volume is still extremely large (41.5\,GB for our high band and low-band data combined).
Since data analysis techniques are still in development, we wish to preserve these full data cubes for the time being.
A factor of two saving can be made by averaging the two polarizations.
A further factor of two can be saved by using 16-bit floats rather than 32-bit floats.
Since any operation (including power spectrum analysis) is going to sum over many values, the minimum precision (approximately 3 decimal digits) is more than adequate provided that care is taken to promote the values to higher precision before operating on them.
Finally, HDF5 supports the compression of each chunk, with the data optionally first being ``shuffled'' (rearranged at the byte level) to improve the compression ratio.
Since the primary beam shape is not rectangular, the compressed size of the dataset can be reduced significantly by setting all pixels beyond a certain point in the primary beam to the same value.
We achieve a final dataset size of $<$18.5GB without averaging polarizations and with minimal blanking.

Code for producing and reading these datafiles is available\footnote{\url{https://github.com/johnsmorgan/imstack}}.
Also provided is parallel code for generating the variability image.
Even on a desktop machine, the running time of this script (about 20 minutes) is negligible compared with time taken to produce the images.

\section*{Acknowledgements}
We would like to thank Elaine Sadler, David Kaplan, and Steven Tingay for their insightful comments on an early draft of this manuscript.
This scientific work makes use of the Murchison Radio-astronomy Observatory, operated by CSIRO. We acknowledge the Wajarri Yamatji people as the traditional owners of the Observatory site. Support for the operation of the MWA is provided by the Australian Government (NCRIS), under a contract to Curtin University administered by Astronomy Australia Limited. We acknowledge the Pawsey Supercomputing Centre which is supported by the Western Australian and Australian Governments.
Parts of this research were conducted by the Australian Research Council Centre of Excellence for All-sky Astrophysics (CAASTRO), through project number CE110001020.
\bibliographystyle{hapj}
\bibliography{refs}
\end{document}